\documentclass[12pt,a4paper]{article}
 
\usepackage{graphicx}
\usepackage{natbib}

\usepackage{my_definitions}
\newcommand{\ur}[2]{#1_{\mbox{\scriptsize #2}}}
\newcommand{\Sk}{\mbox{Sk}}
\newcommand{\Ku}{\mbox{Ku}}

\title{
Lagrangian turbulence in the Adriatic Sea as computed from
drifter data: effects of inhomogeneity and nonstationarity
}
\author{Alberto Maurizi$^{1}$, Annalisa Griffa$^{2,3}$, Pierre-Marie
Poulain$^{4}$\\ Francesco Tampieri$^{1}$\\
\small $^{1}$Consiglio Nazionale Delle Ricerche, ISAC, Bologna, Italy\\
\small $^{2}$Consiglio Nazionale Delle Ricerche, ISMAR, La Spezia, Italy\\
\small $^{3}$RSMAS/MPO, University of Miami, Miami, Florida, USA\\
\small $^{4}$OGS, Trieste, Italy (PIERRE)}

\begin{document}
\maketitle

\begin{abstract}
The properties of mesoscale Lagrangian turbulence in the Adriatic Sea
are studied from a drifter data set spanning 1990-1999, focusing on the
role of inhomogeneity and nonstationarity. A preliminary study is
performed on the dependence of the turbulent velocity statistics on bin
averaging, and a preferential bin scale of $0.25^{\degree}$ is chosen.
Comparison with independent estimates obtained using an optimized spline
technique confirms this choice. Three main regions are identified where
the velocity statistics are approximately homogeneous: the two boundary
currents, West (East) Adriatic Current, WAC (EAC), and the southern
central gyre, CG. The CG region is found to be characterized by
symmetric probability density function of velocity, approximately
exponential autocorrelations and well defined integral quantities such
as diffusivity and time scale. The boundary regions, instead, are
significantly asymmetric with skewness indicating preferential events in
the direction of the mean flow. The autocorrelation in the along mean
flow direction is characterized by two time scales, with a secondary
exponential with slow decay time of $\approx$ 11-12 days particularly
evident in the EAC region. Seasonal partitioning of the data shows that
this secondary scale is especially prominent in the summer-fall season.
Possible physical explanations for the secondary scale are discussed in
terms of low frequency fluctuations of forcings and in terms of mean
flow curvature inducing fluctuations in the particle trajectories.
Consequences of the results for transport modelling in the Adriatic Sea
are discussed.
\end{abstract}

\section{Introduction}\label{sec:intro}

The Adriatic Sea is a semienclosed sub-basin of the Mediterranean Sea
(Fig.1a). It is located in a central geo-political area and it plays an
important role in the maritime commerce. Its circulation has been
studied starting from the first half of the nineteen century
\citep{poulain_etal-c3-2001},
so that its qualitative characteristics have
been known for a long time. A more quantitative knowledge of the
oceanography of the Adriatic Sea, on the other hand, is much more
recent, and due to the systematic studies of the last decades using both
Eulerian and Lagrangian instruments \citep{poulain_etal-c3-2001}.
In particular, a significant contribution to the knowledge of the
surface circulation has been provided by a drifter data set spanning
1990-1999, recently analyzed by \citet{poulain-jms-2001}. These data provide a
significant spatial and temporal coverage, allowing to determine the
properties of the circulation and of its variability.

In \citet{poulain-jms-2001}, the surface drifter data set 1990-1999 has
been analyzed to study the general circulation and its seasonal
variability. The results confirmed the global cyclonic circulation in
the Adriatic Sea seen in earlier studies
\citep{artegiani_etal-jpo-1997}, with closed recirculation cells in the
central and southern regions. Spatial inhomogeneity is found to be
significant not only in the mean flow but also in the Eddy Kinetic
Energy (EKE) pattern, reaching the highest values along the coast in the
southern and central areas, in correspondence to the strong boundary
currents. The analysis also highlights the presence of a marked seasonal
signal, with the coastal currents being more developed in summer and
fall, and the southern recirculating cell being more pronounced in
winter. 

In addition to the information on Eulerian quantities such as mean flow
and EKE, drifter data provide also direct information on Lagrangian
properties such as eddy diffusivity $K$ and Lagrangian time scales $T$,
characterizing the turbulent transport of passive tracers in the basin.
The knowledge of transport and dispersion processes of passive tracers
is of primary importance in order to correctly manage the maritime
activities and the coastal development of the area, especially
considering that the Adriatic is a highly populated basin, with many
different antropic activities such as agriculture, tourism, industry,
fishing and military navigation.

In \citet{poulain-jms-2001}, estimates of $K$ and $T$ have been computed providing
values of $K\approx \unit{2\times10^7}{cm^2~sec^{-1}}$ and $T\approx
\unit{2}{days}$, averaged over the whole basin and over all seasons.
Similar results have been obtained in a previous paper
\citep{falco_etal-jpo-2000},
using a restricted data set spanning 1994-1996. In
\citet{falco_etal-jpo-2000}, the estimated values have also been used as input parameters
for a simple stochastic transport model, and the results have been
compared with data, considering patterns of turbulent transport and
dispersion from isolated sources. The comparison in
\citet{falco_etal-jpo-2000}
is overall satisfactory, even though some differences between data and
model persist, especially concerning first arrival times of tracer
particles at given locations. These differences might be due to various
reasons. One possibility is that the use of global parameters in the
model is not appropriate, since it does not take into account the
statistical inhomogeneity and nonstationarity of the parameter values.
Alternatively, the differences might be due to some inherent properties
of turbulent processes, such as non-gaussianity or presence of multiple
scales in the turbulent field, which are not accounted for in the simple
stochastic model used by \citet{falco_etal-jpo-2000}. These aspects are still
unclear and will be addressed in the present study.

In this paper, we consider the complete data set for the period
1990-1999 as in \citet{poulain-jms-2001}, and we analyze the Lagrangian turbulent
component of the flow, with the goal of
\begin{description}
  \item[]--- identifying the main statistical properties;
  \item[]--- determining the role of inhomogeneity and nonstationarity.
\end{description}
The results will provide indications on suitable transport models for
the area.

Inhomogeneity and nonstationarity for standard Eulerian quantities such
as mean flow and EKE have been fully explored in
\citet{poulain-jms-2001}, while
only preliminary results have been given for the Lagrangian statistics.
Furthermore, the inhomogeneity of probability density function (pdf) shapes (form factors like skewness
and kurtosis) have not been analyzed yet. In this paper, the spatial
dependence of Lagrangian statistics is studied first, dividing the
Adriatic Sea in approximately homogeneous regions. An attempt is then
made to consider the effects on non-stationarity, grouping the data in
seasons, similarly to what done in \citet{poulain-jms-2001} for the Eulerian
statistics.

The paper is organized as follows. A brief overview of the Adriatic Sea
and of previous results on its turbulent properties are provided in
Section 2. In Section 3, information on the drifter data set and on the
methodology used to compute the turbulent statistics are given. The
results of the analysis are presented in Section 4, while a summary and
a discussion of the results are provided in Section 5.

\section{Background}\label{sec:backg}

\subsection {The Adriatic Sea}\label{sec:backg-adri}

\begin{figure}
\includegraphics[width=\textwidth]{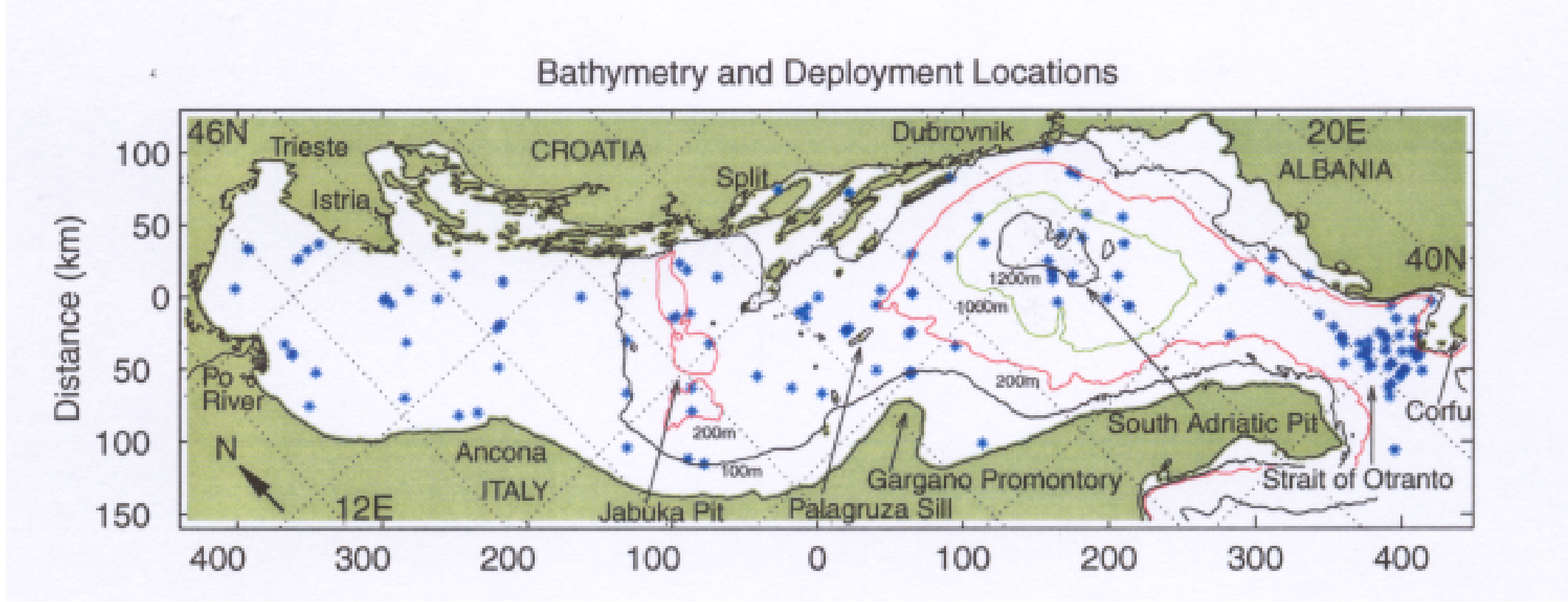}
\caption{}
\label{fig:bati}
\end{figure}

The Adriatic Sea is the northernmost semi-enclosed basin of the
Mediterranean connected to the Ionian Sea at its southern end through
the Strait of Otranto (\figref{fig:bati}). The Adriatic basin, which is
elongated and somewhat rectangular (800 km by 200 km), can be divided
into three distinct regions generally known as the northern, middle and
southern Adriatic \citep{cushman-roisin_etal-2001}. The northern Adriatic
lies on the continental shelf, which slopes gently southwards to a depth
of about 100 m. The middle Adriatic begins where the bottom abruptly
drops from 100 m to over 250 m to form the Mid-Adriatic Pit (also called
Jabuka Pit) and ends at
the Palagruza Sill, where the bottom rises again to approximately 150 m.
Finally, the southern Adriatic, extending from Palagruza Sill to the
Strait of Otranto (780 m deep) is characterized by an abyssal basin
called the South Adriatic Pit, with a maximum depth exceeding 1200 m.
The western coast describes gentle curves, whereas the eastern coast is
characterized by numerous channels and islands of complex topography
(\figref{fig:bati}a).

The winds and freshwater runoff are important forcings of the Adriatic
Sea. The energetic northeasterly bora and the southeasterly sirocco
winds are episodic events that disrupt the weaker but longer-lasting
winds, which exist the rest of the time \citep{poulain_etal-c2-2001};
the Po River in the northern basin provides the largest single
contribution to the freshwater runoff, but there are other rivers and
land runoff with significant discharges \citep{raicich-jms-1996}. Besides
seasonal variations, these forcings are characterized by intense
variability on time scales ranging between a day and a week. 

The Adriatic Sea mean surface flow is globally cyclonic
(\figref{fig:mean}b) due
to its mixed positive-negative estuarine circulation forced by buoyancy
input from the rivers (mainly the Po River) and by strong air-sea fluxes
resulting in loss of buoyancy and dense water formation. The Eastern
Adriatic Current (EAC) flows along the eastern side from the eastern
Strait of Otranto to as far north as the Istrian Peninsula. A return
flow (the WAC) is seen flowing to the southeast along the western coast
\citep{poulain-jms-1999,poulain-jms-2001}. Re-circulation cells embedded in the global
cyclonic pattern are found in the lower northern, the middle and the
southern sub-basins, the latter two being controlled by the topography
of the Mid and South Adriatic Pits, respectively. These main circulation
patterns are constantly perturbed by higher-frequency currents
variations at inertial/tidal and meso- (\eg 10-day time
scale; \citealp{cerovecki_etal-geo-1991}) scales. In particular, the wind stress
is an
important driving mechanism, causing transients currents that can be an
order of magnitude larger than the mean circulation. The corresponding
length scale is 10-20 km, \ie several times the baroclinic radius of
deformation, which in the Adriatic can be as short as 5 km
\citep{cushman-roisin_etal-2001}.

\subsection {Turbulent transport in the Adriatic Sea and previous
drifter studies}\label{sec:backg-transp}

Drifter data are especially suited for transport studies since they move
in good approximation following the motion of water parcels
\citep{niiler_etal-dsr-1995}. As such, drifter data have often been used in the literature
to compute parameters to be used in turbulent transport and dispersion
models \citep{davis-dsr-1991,davis-1994}. In the Adriatic Sea, as mentioned in the
Introduction, turbulent parameters have been previously computed by
\citet{falco_etal-jpo-2000} and \citet{poulain-jms-2001} as global averages over the
basin. A brief overview is given in the following. 

\subsubsection{Models of turbulent transport and parameter definitions}

The transport of passive tracers in the marine environment is usually
regarded as due to advection of the ``mean'' flow, \ie of the large
scale component of the flow $\mathbf{u}(\mathbf{x},t)$, and to
dispersion caused by the ``turbulent'' flow, \ie of the mesoscale and
smaller scale flow. The simplest possible model used to describe these
processes is the advection-diffusion equation, 
\begin{equation}
 \partial C/\partial t + \nabla \cdot(\mathbf{U}C) = \nabla \cdot(\mathbf{K} \nabla C)
 \label{eq:adv-diff}
\end{equation}
where $C$ is the average concentration of a passive tracer,
$\mathbf{U}$ is the mean flow field and $\mathbf{K}$ is the diffusivity tensor
defined as: 
\begin{equation}
 K_{ij}=\int _o^\infty R_{ij}(\tau) d \tau
 \label{eq:diff-coef}
\end{equation}
where $\mathbf{R}(\tau)$ is the Lagrangian autocovariance,
\begin{equation}
 R_{ij}(\tau)=\mean{u_i^\prime(t)u_j^\prime(t+\tau)}
 \label{eq:correl-def}
\end{equation}
with $\mean{\cdot}$ being the ensemble average and
$\mathbf{u}^\prime=\mathbf{u}-\mathbf{U}$ being the turbulent Lagrangian
velocity, \ie the residual velocity following a particle. Note that in
this definition, $\mathbf{R}$ depends only on the time lag $\tau$,
consistently with an homogeneous and steady situation. In fact, non
homogeneous and unsteady flows do not allow for a consistent definition
of the above quantities.

The advection-diffusion equation \eqref{eq:adv-diff} can be correctly
applied only in presence of a clear scale separation between the scale
of diffusion mechanism and the scale of variation of the quantity being
transported \citep{corrsin-advg-1974}. Generalizations of \eqref{eq:adv-diff} are
possible, for example introducing a ``history term'' in
\eqref{eq:adv-diff} that takes
into account the interactions between $\mathbf{U}$ and
$\mathbf{u}^\prime$ \citep[\eg][]{davis-jmr-1987}. Alternatively, a different class
of models can be used, that are easily generalizable and are based on
stochastic ordinary differential equation describing the motion of
single tracer particles \citep[\eg][]{griffa-1996,berloff_etal-jpo-2002}.

A general formulation was given by \citet{thomson-jfm-1987} and further widely
used. The stochastic equations describing the particle state
$\mathbf{z}$ are
\begin{equation}
 \drm z_i = a_i \drm t + b_{ij} \drm W_j
 \label{eq:sde-general}
\end{equation}
where $\drm\mathbf{W}$ is a random increment from a normal distribution
with zero mean and second order moment $\mean{\drm W_i(t)\drm
W_j(s)}=\delta_{ij}\delta(t-s) \drm t$.

\Eqref{eq:adv-diff} can be seen as equivalent to the simplest of these
stochastic models, \ie the pure random walk model, where the particle
state is described by the positions, \ie $\mathbf{z}\equiv\mathbf{x}$
only, which are assumed to be Markovian while the velocity
$\mathbf{u}^\prime$
is a random process with no memory (zero-order model). A more general
model can be obtained considering the particle state defined by its
position and velocity. Thus
$\mathbf{z}\equiv(\mathbf{x},\mathbf{u}^\prime)$
are joint Markovian, so that the turbulent velocity $\mathbf{u}^\prime$
has a finite memory scale, $\mathbf{T}$ (first-order model). In this
case the model can also be applied for times shorter than the
characteristic memory time $T$, in contrast to the zeroth-order model.
If times for which acceleration is significantly correlated is
important, second order models should be used \citep{sawford-blm-1999}. Higher
order models are possible \citep[see, \eg][]{berloff_etal-jpo-2002} but
they require some knowledge on the supposed universal behavior of very
elusive quantities such as time derivatives of tracer acceleration.

For a homogeneous and stationary flow with independent velocity
components, the first-order model can be written for the fluctuating
part $\mathbf{u}^\prime$ for each component and corresponds to the
linear Langevin equation (\ie the Ornstein-Uhlenbeck process, see, \eg
\citealp{risken-1989}):
\begin{eqnarray}
 \drm x_i=(U_i+u_i^\prime)\drm t \label{eq:OU-1}\\
 \drm u_i^\prime=-\frac {u_i^\prime}{T_i} \drm t +
 \sqrt{\frac{2\sigma_i^2}{T_i}} \drm W_i
 \label{eq:OU-2}
\end{eqnarray}
where $\sigma_i^2$ and $T_i$ are the variance and the correlation time
scale of $u_i^\prime$, respectively.

For the model~(\ref{eq:OU-1}--\ref{eq:OU-2}), $u_i^\prime$ is Gaussian and
\begin{equation}
 R_{ii}(\tau)=\sigma_i^2 \exp{(-\frac{\tau}{T_i})},
 \label{eq:exp-corr}
\end{equation}
so that $T_i$
\begin{equation}
 T_i= \frac{1}{\sigma_i^2}\int_o^\infty R_{ii}(\tau)=
 \frac{K_{ii}}{\sigma_i^2}
 \label{eq:correl-time}
\end{equation}
corresponds to the e-folding time scale, or memory scale of $u_i^\prime$.

Description of more complex situations as unsteadiness and
inhomogeneity, as well as non-Gaussian Eulerian velocity field, need the
more general formulation of \citet{thomson-jfm-1987}. An accurate understanding of these
situations is thus necessary in order to properly choose the model to be
applied to describe transport processes to the required level of
accuracy.

\subsubsection{Results from previous studies in the Adriatic Sea}

In \citet{falco_etal-jpo-2000}, the model (\ref{eq:OU-1}--\ref{eq:OU-2}) has been applied using the
drifter data set 1994-1996. The pdf for the meridional and zonal
components of $\mathbf{u}^\prime$, have been computed for the whole dataset
and found to be qualitatively close to Gaussian for small and
intermediate values, while differences appear in the tails. 

For each velocity component, the autocovariance \eqref{eq:correl-def}
has been computed and the parameters $T_i$ and $\sigma_i^2$ have been
estimated: $\sigma_i^2 \approx \unit{100}{cm^2/sec^2}$, $T_i \approx 2$
days. These values have been used also in Lagrangian prediction
studies \citep{castellari_etal-jms-2001} with good results. $R_{ii}(\tau)$
computed in \citet{falco_etal-jpo-2000} appears to be qualitatively similar to
the exponential shape (\eqref{eq:exp-corr}), at least for small $\tau$ whereas it appears
to be different from exponential for time lags $\tau >T_i$, since the
autocovariance tail maintains significantly different from zero.

In \citet{poulain-jms-2001}, estimates of $R_{ii}(\tau)$, $T_i$ and $K_{ii}$ have
been computed using the more extensive data set 1990-1999. A different
method than in \citet{falco_etal-jpo-2000} has been used for the analysis
\citep{davis-dsr-1991}, but the obtained results are qualitatively similar to the
ones of \citet{falco_etal-jpo-2000}. Also in this case, the autocovariance
$R_{ii}(\tau)$ does not converges to zero, resulting in a $K_{ii}$ which
does not asymptote to a constant.

There might be various reasons for the observed tails in the
autocovariances and in the pdf. First of all, they might be an effect
of poorly resolved shears in the mean flow $\mathbf{U}$. This aspect has
been partially investigated in \citet{falco_etal-jpo-2000} and \citet{poulain-jms-2001}
using various techniques to compute $\mathbf{U}(\mathbf{x})$. Another
possible explanation is related to unresolved inhomogeneity and
nonstationarity in the turbulent flow. Since the estimates of the pdf
and autocovariances are global, over the whole basin and over the whole
time period, they might be putting together different properties from
different regions in space and time, resulting in tails. Finally, the
tails might be due to inherent properties of the turbulent field, which
might be different from the simple picture of an Eulerian Gaussian pdf
and an exponential Lagrangian correlation for $\mathbf{u}^\prime$.

In this paper, these open questions are addressed.
A careful examination of the dependence of turbulent statistics
on the mean flow
$\mathbf{U}$ estimation is performed. Possible dependence on spatial
inhomogeneity is studied, partitioning the domain in approximatively
homogeneous regions. Finally, an attempt to resolve seasonal time dependence
is performed.

\section{Data and methods}\label{sec:data}

\subsection{The drifter data set}\label{sec:data-set-descrip}

As part of various scientific and military programs, surface 
drifters were launched in the Adriatic in order to measure 
the temperature and currents near the surface. Most of 
the drifters were of the CODE-type and followed the currents 
in the first meter of water with an accuracy of a 
few cm/s \citep{poulain_etal-1998,poulain-jms-1999}. They 
were tracked by, and relayed SST data to, the Argos system onboard 
the NOAA satellites. More details on the drifter design, 
the drifter data and the data processing can be found in 
\citet{poulain_etal-2003}. Surface velocities were calculated from 
the low-pass filtered drifter position data and do not include 
tidal/inertial components. The Adriatic drifter database 
includes the data of 201 drifters spanning the time period 
between 1 August 1990 and 31 July 1999. It contains time 
series of latitude, longitude, zonal and meridional 
velocity components and sea surface temperature, all sampled 
at 6-h intervals. Due to their short operating lives (half life 
of about 40 days), the drifter data distribution is very sensitive 
to the specific locations and times of drifter deployments. 
The maximum data density occurs in the southern Adriatic 
and in the Strait of Otranto. Most of the observations 
correspond to the years 1995-1999.

\subsection{Statistical estimate of the mean flow: averaging scales}\label{sec:stat}

Estimating the mean flow $\mathbf{U}(\mathbf{x},t)$ is of crucial
importance for the identification of the turbulent component
$\mathbf{u}^\prime$, since $\mathbf{u}^\prime$ is computed as the velocity residual
following trajectories. If the space and time scales of
$\mathbf{U}(\mathbf{x},t)$ are not
correctly evaluated, they can seriously contaminate the statistics of
$\mathbf{u}^\prime$. Particularly delicate is the identification of the
space scales of the mean shears in $\mathbf{U}$,
since they can be relatively small (of the same order as the scales
of turbulent mesoscale variability), and, if not resolved, they
can result in persistent tails in
the autocovariances and spuriously high
estimates of turbulent dispersion \citep[\eg][]{bauer_etal-jgr-1998}. Identifying a correct averaging
scale $\ur{L}{a}$ for estimating $\mathbf{U}$ is therefore a 
very important issue
for estimating the $\mathbf{u}^\prime$ statistics.

Various methods can be used to estimate $\mathbf{U}$. Here we consider
two methods: the classic methods of bin averaging and a method based on
optimized bicubic spline interpolation (Inoue 1986, Bauer et al., 1998).
Results from the two methods are compared, in order to test their
robustness. The results from bin averaging are discusses first, since
the method is simpler and it allows for a more straightforward analysis
of the impact of the averaging scales on the estimates. 

For the bin averaging method, $\ur{L}{a}$ simply corresponds to the
bin size. In principle, given a sufficiently high number of data, an
appropriate averaging scale $\hat{\ur{L}{a}}$ can be identified such that the
mean flow shear is well resolved. The $\mathbf{u}^\prime$ and
$\mathbf{U}$ statistics
are expected to be independent on ${\ur{L}{a}}$ for ${\ur{L}{a}} <
\hat{\ur{L}{a}}$. In
practical applications, though, the number of data is limited and the
averaging scale is often chosen as a compromise between the high
resolution, necessary to resolve the mean shear, and the data density
per bin, necessary to ensure significant estimates. In practice, then,
$\ur{L}{a}$ is often chosen as ${\ur{L}{a}} > \hat{\ur{L}{a}}$ and the asymptotic
independency of the statistics on $\ur{L}{a}$ is not reached. 

\citet{poulain-jms-2001} tested the dependence of the mean and eddy kinetic
energy, MKE and EKE, on the bin averaging scale $\ur{L}{a}$ for the 1990-1999
data set. Circular, overlapping bins with radius varying between 400
and 12.5 km were considered. It was found that, in the considered range,
EKE and MKE (computed over the whole basin) do not converge toward a
constant at decreasing $\ur{L}{a}$. A similar calculation is repeated
here (\figref{fig:binning}),
considering some modifications. First of all, we consider
square bins nonoverlapping, to facilitate the computation of turbulent
statistics, such as $R(\tau)$, which involve particle tracking. Also the
EKE and MKE estimates are computed considering only ``significant''
bins, \ie bins with more than 10 independent data, $n_{bi} > 10$, where
$n_{bi}$ is computed resampling each trajectory with a period $T$ = 2
days, on the basis of previous results from \citet{falco_etal-jpo-2000} and
\citet{poulain-jms-2001}. Finally, the values of
EKE and MKE are displayed in \figref{fig:binning} together with a parameter,
$N_{\ur{L}{a}}/\ur{N}{tot}$, providing information on the statistical significance
of the results at a given $\ur{L}{a}$. $N_{\ur{L}{a}}/\ur{N}{tot}$, in fact is the
fraction of data actually used in the estimates (\ie belonging to the
significant bins) over the total amount of data in the basin
$\ur{N}{tot}$.

The behavior of EKE and MKE in \figref{fig:binning} is qualitatively similar to what
shown in \citet{poulain-jms-2001}, even though the considered range is slightly
different and reaches lower values of $\ur{L}{a}$ (bin sizes vary between
$1^{\degree}$ and $0.05^{\degree}$). The values of EKE and MKE do not appear to
converge at small $\ur{L}{a}$, but the interesting point is that they tend to
vary significantly for $\ur{L}{a} < 0.25^{\degree}$, \ie in correspondence to the
drastic decrease of $N_{\ur{L}{a}}/\ur{N}{tot}$. This suggests that the strong lack
of saturation at small scales is mainly due to the fact that
increasingly fewer bins are significant and therefore the statistics
themseplves become meaningless.
These considerations suggest that the ``optimal'' scale $\ur{L}{a}$, given the
available number of data $\ur{N}{tot}$ is of the order of $0.25^{\degree}$, since
it allows for the highest shear resolution still maintain a significant
number of data ($\approx 80 \%$). This choice is in agreement with
previous results by \citet{falco_etal-jpo-2000} and
\citet{poulain-jms-1999}.

The binned mean field $\mathbf{U}$ obtained with the $0.25^{\degree}$ bin
(\ie between 19
and 28 km) is shown in \figref{fig:mean-flow}. As it can be seen, it is qualitatively similar to
the $\mathbf{U}$ field obtained by \citet[Fig. 1b]{poulain-jms-2001} with a 20 km circular 
bin average.

 As a further check on the binned results and on the $\ur{L}{a}$ choice,
a comparison is performed with results obtained using the spline method
\citep{bauer_etal-jgr-1998,bauer_etal-jgr-2002}. This method, previously applied by Falco et al., 
(2000) to the 1994-1996 data set, is based on a bicubic spline
interpolation \citep{inoue-geo-1986} whose parameters are optimized in order to guarantee minimum
energy in the fluctuation field $\mathbf{u}^\prime$ at low frequencies. 
Notice that, with respect to the binning average technique,
the spline method has the advantage that the estimated
$\mathbf{U}(\mathbf{x})$ 
is a smooth function of space. As a consequence,
the values of the turbulent residuals $\mathbf{u}^\prime$ can be computed subtracting
the exact values of $\mathbf{U}$ along trajectories, instead than
considering discrete average values inside each bin. In other words, the
spline technique allows for a better resolution of the shear
inside the bins. 

Details on the choice of the spline parameters are given in Appendix.
The resulting statistics are compared with the binned
results in \figref{fig:binning}. The turbulent residual
$\mathbf{u}^\prime$ has been computed subtracting the splined $\mathbf{U}$, and
the associated EKE have been calculated
as function of $\ur{L}{a}$. The EKE
dependence on size for very small scales is due to the fact that EKE is computed as an average over
significant bins and the number of bins decreases for small
$\ur{L}{a}$.
In the case of the binned $\mathbf{U}$ described before, instead,
also the estimates of $\mathbf{U}$ and $\mathbf{u}^\prime$ inside each grid change and
deteriorate as
$\ur{L}{a}$ decreases. As a consequence, it is not surprising that the EKE values
change much less in the splined case with respect to the binned case.
Notice that the splined EKE values are very similar
to the binned ones for bins in the range between $0.35^{\degree}-0.25^{\degree}$.
This provides support to the choice of
$\ur{L}{a}=0.25^{\degree}$. Also, a direct comparison between the splined (not shown) and
binned $\mathbf{U}$ fields show a great similarity, 
as already noticed also in the case of Falco et
al. (2000). 

In conclusion, the spline analysis confirms that the choice of 
$\ur{L}{a} = 0.25^{\degree}$ is appropriate. $\ur{L}{a} = 0.25^{\degree}$, in fact, provides robust
estimates while resolving 
the important spatial variations of the mean flow and averaging the mesoscale.

\subsection{Homogeneous regions for turbulence statistics}\label{sec:homo}

We are
interested in identifying regions where the $\mathbf{u}^\prime$ statistics can be
considered approximately homogeneous, so that the main turbulent 
properties can be meaningfully studied. In a number of 
studies in various oceans and for various data sets
\citep{swenson_etal-jgr-1996,bauer_etal-jgr-2002,veneziani_etal-jpo-2003},
``homogeneous'' regions have been identified as regions
with consistent dynamical and statistical properties.
A first qualitative identification of consistent dynamical regions in the
Adriatic Sea can be made based on the literature and on the knowledge of
the mean flow and of the topographic structures (\figref{fig:bati}). 

 First of all, two boundary current regions can be identified, along the
eastern coast (Eastern Adriatic Current, EAC) and western coast
(Western Adriatic Current, WAC).
These regions are characterized by strong mean flows and well organized
current structure. A third region can be identified with the central area of
the cyclonic gyre in the south/central Adriatic (Central Gyre, CG).
This region is characterized by a deep topography (especially in the southern
part) and by a weaker mean flow structure.
Finally, the northern part of the basin, characterized by shallow depth ($<
\unit{50}{m}$), could be considered as a forth region (Northern Region, NR). 
With respect to the other regions, though, NR appears less dynamically
homogeneous, given that the western side is heavily dominated by
buoyancy forcing related to the Po river discharge, while the eastern part
is more directly influenced by wind forcing. Also, NR has a lower data
density with respect to the other regions \citep{poulain-jms-2001}. For these
reasons, in the following we will focus on EAC, WAC and CG. A complete
analysis of NR will be performed in future works, when more data will be
available.

 As a second step, a quantitative definition of the boundaries between
regions must be provided. Here we propose to use as a main parameter to
discriminate between regions the relative turbulence intensity
$\gamma=\sqrt{\mbox{EKE}/\mbox{MKE}}$. The parameter $\gamma$ is
expected to vary from $\gamma <1$ in
the boundary current regions dominated by the mean flow,
to $\gamma >1$ in the the central gyre region dominated by fluctuations.

A scatterplot of $\gamma$ versus $\sqrt{\mbox{MKE}}$ is shown in
\figref{fig:partition}.
Two well defined regimes can be seen, with $\gamma <1$ and $\gamma >1$
respectively. The two regimes are separated
by $\sqrt{\mbox{MKE}} \approx \unit{6-7}{cm~sec^{-1}}$.
Based on this result, we use the (conservative) value $\sqrt{\mbox{MKE}} = \unit{8}{cm~sec^{-1}}$
to discriminate between regions. The resulting partition is shown in
\figref{fig:mean-flow}. As it can be seen, the regions (indicated by the different colors
of the mean flow arrows) appear well defined, indicating that the
criterium is consistent. The WAC region reaches the northern part of
the basin, up to $\approx 44^{\degree} N$, because of the influence of the Po
discharge on the boundary current. The EAC region, on the other hand, is
directly influenced by the Ionian exchange through the Otranto Strait
and it is limited to the south/central part of the basin, connected to
the cyclonic gyre. The CG region appears well defined in the center of
the two recirculating cells in the southern and central basin.

It is interesting to compare the regions defined in
\figref{fig:mean-flow} with the
pattern of EKE computed by \citet[Fig. 4d]{poulain-jms-2001}. The two
boundary regions EAC and WAC, even though characterized by
$\mbox{EKE}/\mbox{MKE} < 1$, correspond to regions of high EKE values,
EKE $> \unit{100}{cm^2~sec^{-2}}$. The CG region, instead is characterized by
low EKE values, approximately constant in space. The three regions,
then, appear to be quasi-homogeneous in terms of EKE values, confirming
the validity of the partition. The northern region NR, on the other
hand, shows more pronounced gradients of EKE, with EKE $>\unit{100}{cm^2
sec^{-2}}$ close to the Po delta, EKE $\approx \unit{50}{cm^2~sec^{-2}}$ in
the central part and lower values in the remaining parts. This confirms
the fact that NR cannot be considered a well defined homogeneous region
as the other three, and it will have to be treated with care in the
future, with a more extensive data set.

The main diagnostics presented hereafter and computed for each region are:

 \begin{description}
\item [] Characterization of the $\mathbf{u}^\prime$ pdf. Values of skewness and
kurtosis will be evaluated and compared with standard Gaussian values

\item [] Estimation of $\mathbf{u}^\prime$ autocorrelations,
$\rho_{i}(\tau)=R_{ii}(\tau)/{\sigma_i^2}$. They will be qualitatively compared to
the exponential shape (7) and estimates of e-folding time
scales will be performed.
Estimation of integral quantities such as diffusivity $\mathbf{K}$ from
\eqref{eq:diff-coef}
and integral time scale $\mathbf{T}$ from \eqref{eq:correl-time} will also be performed.


 \end{description}

These quantities will be first computed as averages over the whole
time period, and then an attempt to separate the data seasonally will be
performed.

Since all the quantities are expressed as vector components, the choice
of the coordinate system is expected to play a role in the presentation
of the results. It is expected that the mean flow (when significant)
could influence turbulent features resulting in an anisotropy of
statistics. Thus, in the following, we consider primarily a ``natural''
coordinate system, which describes the main properties more clearly.
The natural Cartesian system is obtained rotating locally along the mean
flow axes. The components of a quantity $\mathbf{Q}$ in that system are the
streamwise componente $Q_{\parallel}$ and the across-stream componente $Q_{\perp}$.

\section{Results} 

\subsection{Statistics in the homogeneous regions} 

Here the statistics of $\mathbf{u}^\prime$ in the three regions identified in
Section 3.3 are computed averaging over the whole time period, \ie
assuming stationarity over the 9 years of measurements. In all cases,
$\mathbf{u}^\prime$ is computed as residual velocity with respect to the
$0.25^{\degree} \times 0.25^{\degree}$ binned mean flow, as explained in
Section 3.2. In some selected cases, results from other bin sizes and
from the spline method are considered as well, in order to further test
the influence of the $\mathbf{U}$ estimation on the results. As in
Section \ref{sec:data},
the statistics are computed only in the significant bins,
$\ur{n}{b}>10$. Also, data points with velocities higher than 6 times the
standard deviations have been removed. They represent an ensemble of
isolated events that account for 10 data points in total, distributed
over 4 drifters. While they do not significantly affect the second
order statistics, they are found to affect higher order moments such as
skewness and kurtosis.

\subsubsection{Characterization of the velocity pdf}

The pdf of $\mathbf{u}^\prime$ is computed normalizing the velocity
locally, using the variance $\sigma^2_b$ computed in each bin
\citep{bracco_etal-jpo-2000}. This is done in order to remove possible
residual inhomogeneities inside the regions. The pdfs are characterized
by the skewness $\Sk =\mean{{u^\prime}^3}/\sigma^3$ and the kurtosis
$\Ku =\mean{{u^\prime}^4}/\sigma^4$.  Here we follow the results of
\citet{lenschow_etal-jaot-1994}, which provide error estimates for
specific processes at different degrees of non-gaussianity as function of
the total number of independent data $N_i$.  In the range of our data
(\Tabref{tab:1}), the mean square errors of $\Sk$ and $\Ku$ from
\citet{lenschow_etal-jaot-1994} appear to be $(\delta \Sk)^2 \approx
10/N_i$, $(\delta \Ku)^2 \approx 330/N_i$. Notice that these values can
be considered only indicative, since they are obtained for a specific
process.

Before going into the details of the results and discussing them from a
physical point of view, a preliminary statistical analysis is carried
out to test the dependence of the higher moments $\mathbf{Sk}$ and
$\mathbf{Ku}$ from the
bin size, similarly to what done in Section 3.2 for the lower order
moments. In \figref{fig:binning-S}a,b, estimates of $\Sk$ and $\Ku$ computed over the whole
basin (in Cartesian coordinate) are shown, at varying bin size from
$1^{\degree}$ to $0.2^{\degree}$ (smaller bins are not considered given the small number
of independent data, see \figref{fig:binning}). Given that the total number of
independent data is of the order of $N_i \approx 4000$, the error
estimates from \citet{lenschow_etal-jaot-1994} suggest $\sqrt{(\delta \Sk)^2} \approx
0.05$, $\sqrt{(\delta \Ku)^2} \approx 0.25$. As it can be seen, the
values of $\Sk$ and $\Ku$ do not change significantly in the range
$0.5^{\degree}-0.25^{\degree}$. Values of $\Sk$ and $\Ku$
have also been computed using splined estimates (not shown), and they are found to
fall in the same range. These results confirm the choice of the
$0.25^{\degree}$ binning of Section 3.2. Notice that, since $\Sk$ and $\Ku$ in
\figref{fig:binning-S}a,b are computed averaging over different dynamical regions, their
values do not have a straightforward physical interpretation. We will
come back on this point in the following, after analyzing the specific
regions.

The pdfs for the three regions computed with the $0.25^{\degree}$ binning are
shown in \figref{fig:pdf-EAC}a,b,c in natural coordinates, while the $\Sk$ and $\Ku$ values are
summarized in \Tabref{tab:1}. For
each region, $N_i \approx 1000$, so that $\sqrt{(\delta \Sk)^2} \approx
0.1$, $\sqrt{(\delta \Ku)^2} \approx 0.5$. Furthermore, a quantitative
test on the deviation from gaussianity has been performed using the
Kolmogorov-Smirnov test \citep{priestly-1981,num_recipes}. Notice that the K-S test is known to be
mostly sensitive to the distribution mode (\ie to the presence of
asymmetry, or equivalently to $\Sk$ being different from zero),
while it can be quite insensitive to the existence of tails in the
distribution (large $\Ku$). More sophisticated tests should be used to
guarantee sensitivity to the tails.

Let's start discussing the Eastern boundary region, EAC. The $\Sk$ 
is positive and significant in the along
component ($\Sk_{\parallel} \approx 0.48$), while it is only marginally different
from zero in the cross component ($\Sk_{\perp} \approx -0.14$). Positive
skewness indicates that the probability of finding high positive values
of $u^\prime_{\parallel}$ is higher than the probability of negative high values,
(while the opposite is true for small values). This is also shown by the
pdf shape (\figref{fig:pdf-EAC}a). Physically, this indicates the existence of an
anisotropy in the current, with the fluctuations being more energetic in
the direction of the mean flow. This asymmetry is not surprising, given
the existence of a privileged direction in the mean. This fact
has long been recognized in boundary layer flows \citep[\eg][]{durst_etal-1987}. The cross
component, on the other hand, does not have a privileged direction and
its $\Sk$ is much smaller, as shown also by the pdf shape.
The values of the kurtosis $\Ku$ are around 4 for both components, 
indicating high probability for energetic events. This
is clear also from the high tails in the pdf.

The K-S statistics computed for the pdfs of \figref{fig:pdf-EAC}a are $\alpha=0.012$
for $u^\prime_{\parallel}$ indicating rejection of the null hypothesis (that the
distribution is Gaussian) at the $95\%$ confidence level. For the cross
component $u^\prime_{\perp}$, instead, $\alpha=0.09$ so that the null
hypothesis cannot be rejected. It is worth noting that the estimates of
$\alpha$ depend on the number of independent data $N_i$, which in turn
depends on $T$. Here, $T$ is assumed $T=$ 2 days. For the cross
component, this is probably an overestimate (as it will be shown in the
following, see \figref{fig:corr-u}b), and $T=$ 1 day is probably a better assumption.
Even if computed with $T=$ 1 days, $\alpha=0.04$ for $u^\prime_{\perp}$,
suggesting that the Gaussian hypothesis can be only marginally rejected.

The results for the western boundary region WAC are qualitatively
similar to the ones foe EAC. The $\Sk$ values in natural coordinates are
$\Sk_{\parallel} \approx 0.52$ and $\Sk_{\perp} \approx 0.09$, suggesting the same
along current anisotropy found in EAC. 
Notice that the total value of $\Sk$ computed over the whole basin
(\figref{fig:binning-S}a) is
approximately zero, because the two contributions from the two boundary
currents nearly cancel each other when computed in fixed cartesian
coordinates.

Also the structure of the pdfs (\figref{fig:pdf-EAC}b) are qualitatively similar to
the EAC ones, exhibiting a clear asymmetry and high tails, especially
for $u^\prime_{\parallel}$. The K-S statistics are $\alpha=0.027$ for
$u^\prime_{\parallel}$,
suggesting a significant deviation from gaussianity. For $u^\prime_{\perp}$,
on the other hand, $\alpha=0.4$ ($\alpha=0.097$ for $T=$ 1 day), which
is not significantly different from Gaussian.

The central region, CG, has lower values of $\Sk$ in both components 
($0.16$ and $-0.02$ respectively). This is shown also by the
pdf patterns (\figref{fig:pdf-EAC}c), which are more symmetric than for EAC and WAC.
This is not surprising given that the mean flow is weaker in CG, so that
there is no privileged direction. The tails, on the other hand, are
high also in CG, as shown by the $\Ku$ values that are in the same range
(and actually slightly higher) than for EAC and WAC. The K-S statistics
do not show a significant deviation from gaussianity in any of the two
components, $\alpha=0.44$ for $u^\prime_{\parallel}$ and $\alpha=0.33$ for
$u^\prime_{\perp}$ ($\alpha=0.058$ for $T=$ 1 day). This is due to the fact
that the K-S test is mostly sensitive to the mode, as explained above.

In summary, the turbulent component along the mean flow is significantly
non Gaussian and, in particular, asymmetric in both boundary currents.
The strong mean flow determines the existence of a privileged direction,
resulting in anisotropy of the fluctuation, with more energetic events
in the direction of the mean. The central gyre region and the cross
component of the boundary currents, do not appear significantly skewed.
For all regions and all components, though, the kurtosis is higher than
3 consistently with other recent findings \citep{bracco_etal-jpo-2000}.
indicating the likelihood of high energy events.

\subsubsection{Autocorrelations of $\mathbf{u}^\prime$}

The autocorrelations in natural components are shown in
\figref{fig:corr-u}a,b
The along component results $\rho_{\parallel}(\tau)$, are shown in
\figref{fig:corr-u}a
for the 3 regions and for the whole basin. Errorbars are computed as
$1/N$ where $N$ is the number of independent data for each time lag
$t$. The autocorrelation for the whole
basin shows
two different regimes with approximately exponential behavior.
The nature of this shape can be better investigated considering the
three homogeneous regions separately. For small lags exponential
behavior is evident in all the three regions, with a slightly different
e-folding time scales: $\ur{\tau}{exp} \approx 1.8$ days for EAC and WAC
and $\approx 1.1$ days for CG. The above values were computed fitting
the exponential function on the first few time lags. This is consistent with the fact that
$\ur{\tau}{exp}$ is representative of fluctuations due to processes
such as internal instabilities and direct wind forcing, which are
expected to be different in the boundary currents and in the gyre center.
At longer lags $\tau >\unit{3-4}{days}$, the behavior in the three regions
become even more distinctively different. In region EAC, a clear change
of slope occurs, indicating that $\rho_{\parallel}$ can be characterized by a
secondary exponential behavior with a slower decay time of $\approx$
11-12 days. Only a hint of this secondary scale is present in WAC,
while there is no sign of it in CG. The behavior of the basin average
$\rho_{\parallel}$, then, appears to be determined mostly by the EAC region. 

In contrast to the along component behavior, the cross component,
${\rho_{\perp}}$, (\figref{fig:corr-u}b) appears characterized by a fast decay in all
three region as well as in the basin average ($\ur{\tau}{exp} \approx
0.5-0.7$ days), with a significative loss of correlation for time lags
less than 1 day. This can be qualitatively understood
considering as a reference the behavior of parallel and transverse
Eulerian correlations in homogeneous isotropic turbulence
\citep{batchelor-1970}.
It indicates that the turbulent fluctuations, linked to mean flow
instabilities, tend to develop structures oriented along the mean
current. As a consequence, the cross mean flow dispersion is found to be
very fast and primarily dominated by a diffusive regime, while the along
mean dispersion tends to be slower and dominated by more persistent
coherent structures. This result suggest that a correlation time of 2
days (as estimated in \citealp{poulain-jms-2001,falco_etal-jpo-2000}) is actually a measure of mixed
properties.

In summary, the results show that the eastern boundary region EAC is
intrinsically different from the center gyre region CG and also
partially different from the western boundary region WAC. While CG (and
partially WAC) are characterized by a single scale of the order of 1-2
day, EAC is clearly characterized by 2 different time scales, a fast
one (order of 1-2 days) and a significantly longer one (order of 11-12
days). The physical reasons behind this two-scale behavior is not
completely understood yet, and some possible hypotheses are discussed
below.

Falco et al (2000) suggested that the observed autocorrelation tails
could be due to a specific late summer 1995 event sampled by a few
drifters launched in the Strait of Otranto. In order to test this
hypothesis, we have removed this specific subset of drifters and
re-computed ${\mathbf{\rho_{\parallel}}}$. The results (not shown) do not show
significant differences and the 2 scales are still evident.

A possible hypothesis is that the 2 scales are due to different
dynamical processes co-existing in the system. The short time scale
appears almost certainly related to mesoscale instability and
wind-driven synoptic processes, while the longer time scale might be
related to low frequency fluctuations in the current, due for instance
to changes in wind regimes or to inflow pulses through the Strait of Otranto
This is suggested by the presence of a 10 day period fluctuation
in Eulerian currentmeter records \citep{poulain-jms-1999}.
Finally, another possibility is that the longer time scale is
related to the spatial structure of the mean flow, namely its
curvatures. Such curvature appears more pronounced and consistently present
in the 
circulation pattern of the EAC than in the WAC, in
agreement with the fact that the the secondary scale is more evident in
EAC. At this point, not enough data are available to quantitatively
test these hypotheses and clearly single out one of them.

\subsubsection{Estimates of $\mathbf{K}$ and $\mathbf{T}$ parameters}

From the autocorrelations of \figref{fig:corr-u}a,b, the components of
the diffusivity and integral time scale \eqref{eq:diff-coef} and
\eqref{eq:correl-time} can be computed by integration. $\mathbf{T}$ and
$\mathbf{K}$ are input parameters for models, and are therefore of great
importance in practical applications.  Estimates of the natural
components of $\mathbf{T}$, $T_{\parallel}(t)$ and $T_{\perp}(t)$, are
shown in \figref{fig:T-u}a,b for the three regions and for $t<$ 10 days.
The behavior of the $\mathbf{K}$ components is the same as for
$\mathbf{T}$, since for each component $T(t)=K(t)/\sigma^2$ (8). The
values of $T_{\parallel}(t)$, $T_{\perp}(t)$. $K_{\parallel}(t)$.
$K_{\perp}(t)$ at the end of the integration, at $t=\unit{10}{days}$,
are reported in \Tabref{tab:2}.

The along component $T_{\parallel}(t)$ (\figref{fig:T-u}a) shows a
significantly different behavior in the three regions. In CG,
$T_{\parallel}(t)$ converges toward a constant, so that the asymptotic
value is well defined, $T_{\parallel} \approx$ 1.2 day. This
approximately corresponds to the estimate of $\ur{\tau}{exp} \approx
0.8$ from \figref{fig:corr-u}a. In EAC, instead, $T_{\parallel}$ is not
well defined, given that $T_{\parallel}(t)$ keeps increasing, reaching a
value of $\approx$ 2.7 days at $t=$ 10 days, significantly higher than
$\ur{\tau}{exp} \approx \unit{1.4}{days}$.  Finally, WAC shows an
intermediate behavior, with $T_{\parallel}(t)$ growing slowly and
reaching a value of $\approx \unit{1.9}{days}$, slightly higher than
$\ur{\tau}{exp} \approx \unit{1.4}{days}$. These results are consistent with
the shape of $\rho_{\parallel}$ (\figref{fig:corr-u}a) in the three
regions. The values of $K_{\parallel}$ (10 days) (\Tabref{tab:2}) range
between $\unit{0.7\times 10^6}{cm^2~sec^{-1}}$ and $\unit{3.8\times
10^7}{cm^2~sec^{-1}}$ showing a marked variability because of the
different EKE in the three regions.

The cross component $T_{\perp}(t)$ (\figref{fig:T-u}b) shows little
variability in all the three regions, again in keep with the
$\rho_{\perp}$ results (\figref{fig:corr-u}b). In all the regions,
$T_{\perp}(t)$ converges toward a constant value of $T_{\perp} \approx
\unit{0.52-0.78}{days}$, in the same range as the $\ur{\tau}{exp}$
values. More in details, notice that in WAC $T_{\perp}(t)$ tends to
decrease slightly, possibly in correspondence to saturation phenomena
due to boundary effects. The values of $K_{\perp}$ (10 days) in
\Tabref{tab:2} range between $\unit{1.4\times 10^6}{cm^2~sec^{-1}}$ and
$\unit{3.1\times 10^6}{cm^2~sec^{-1}}$.

In summary, the results show that the cross components $T_{\perp}$ and
$K_{\perp}$ are well defined in the three regions, with $T_{\perp}$
approximately corresponding to $\ur{\tau}{exp}$. The along components
$T_{\parallel}$ and $K_{\parallel}$, instead, are well defined only in CG,
while in the boundary regions and especially in EAC, there is no
convergence to an asymptotic value.

The observed values are quite consistent with the averages
reported in \citet{poulain-jms-2001}. Remarkably, in that paper, the
strong inhomogeneity and anisotropy of the flow in the basin was
outlined, noting that the estimates of the time scales for the along
flow components in the boundary currents is significantly larger than
the one related to the central gyre. 

\subsection{Seasonal dependence} 

As an attempt to consider the effects of non-stationarity, a time
partition of the data is performed grouping them in seasons. The data
are not sufficient to resolve space and time dependence together, since
the $\mathbf{u}^\prime$ statistics are quite sensitive involving higher
moments and time lagged quantities. For this reason, averaging is
computed over the whole basin and two main extended seasons are
considered. Based on preliminary tests and on previous results by
\citet{poulain-jms-2001}, the following time partition is chosen: a
summer-fall season, spanning July to December, and a winter-spring
season, spanning January to June.
 
As in Section 4.1, $\mathbf{u}^\prime$ is computed as residual velocity
with respect to the $0.25^{\degree} \times 0.25^{\degree}$ binned mean
flow $\mathbf{U}$.  Mean flow estimates in the 2 seasons are shown in
\figref{fig:mean-season-u}a,b.  As discussed in \citet{poulain-jms-2001},
during summer-fall the mean circulation appears more energetic and
characterized by enhanced boundary currents.  During the winter-spring
season, instead, mean currents are generally weaker and the southern
recirculating gyre is enhanced. 

The $\mathbf{u}^\prime$ statistics during the 2 seasons are
characterized by the autocorrelation functions shown in
\figref{fig:corr-season-u}a,b. The
along component $\rho_{\parallel}(\tau)$
(\figref{fig:corr-season-u}a) has a distinctively
different behavior in the 2 seasons. In summer-fall, the overall
behavior is similar to the one obtained averaging over the whole period
(\figref{fig:corr-u}a). Two regimes can be seen, one approximately exponential at
small lags, and a secondary one at longer lags, $\tau >3 days$, with
significantly slower decay. This secondary regime is not observed in
winter-spring. As for the cross component $\rho_{\perp}(\tau)$
(\figref{fig:corr-season-u}b), both seasons appears characterized by a fast decay, as in the
averages over the whole period (\figref{fig:corr-season-u}b).

Various possible explanations for the longer time scale in
$\rho_{\parallel}(\tau)$ have been discussed in Section 4.1 for the
whole time average.  They include low-frequency forcing and current
fluctuations, as well as the effects of the mean flow curvature in the
boundary currents.  The summer-fall intensification of
\figref{fig:corr-season-u}a,b does not
rule out any of these explanations, given that the strength and
variability of the boundary currents are intensified especially in the
fall.

\subsection{Summary and concluding remarks} 

In this paper, the properties of the Lagrangian mesoscale turbulence
$\mathbf{u}^\prime$ in the Adriatic Sea (1990-1999) are investigated, with
special case to give a quantitative estimate of spatial inhomogeneity
and nonstationarity.

The turbulent field $\mathbf{u}^\prime$ is estimated as the residual velocity
with respect to the mean flow $\mathbf{U}$, computed from the data using
the bin averaging technique. In a preliminary investigation, the
dependence of $\mathbf{u}^\prime$ on the bin size $\ur{L}{a}$ is studied and a
preferential scale $\ur{L}{a}= 0.25^{\degree}$ is chosen. This scale allows
for the highest mean shear resolution still maintaining a significant
amount of data ($\approx 80 \%$). Values of higher moments such as
skewness $\Sk$ and kurtosis $Kr$ are found to be approximately constant
in the $\ur{L}{a}$ range around $0.25^{\degree}$. Further support to the
choice $\ur{L}{a}= 0.25^{\degree}$ is given by the comparison with results
obtained with independent estimates of $\mathbf{U}$ based on an optimized
spline technique \citep{bauer_etal-jgr-1998,bauer_etal-jgr-2002}.

The effects of inhomogeneity and stationarity are studied separately,
because there are not enough data to perform a simultaneous partition in
space and time. The spatial dependence is studied first, partitioning
the basin in approximately homogeneous regions and averaging over the
whole time period. The effects of nonstationarity are then considered,
partitioning the data seasonally, and averaging over the whole basin.

Three main regions where the $\mathbf{u}^\prime$ statistics can be considered
approximately homogeneous are identified. They correspond to the two
(eastern EAC, and western WAC) boundary current regions, characterized
by both strong mean flow and high kinetic energy ($\sqrt{EKE/MKE}<1)$),
and the central gyre region CG in the southern and central basin,
characterized by weak mean current and low eddy kinetic energy
($\sqrt{EKE/MKE}>1$). The northern region is not included in the study
because, in addition to have a lower data density with respect to the
other regions, it appears less dynamically and statistically
homogeneous.

The properties of $\mathbf{u}^\prime$ in the three regions are studied
considering pdfs, autocorrelations and integral quantities such as
diffusivity and integral time scales. Natural coordinates, oriented
along the mean flow direction, are used, since they allow to better
highlight the dynamical properties of the flow.

The pdfs results indicate that the CG region is in good approximation
isotropic with high kurtosis values, while the along components of the
boundary regions EAC and WAC show significant asymmetry (positive
skewness). This is related to energetic events occurring preferentially
in the same direction as the mean flow. Both boundary regions appear
significantly non-Gaussian, while the Gaussian hypothesis cannot be
rejected in the CG region.

Both components of the autocorrelation are approximately exponential in
CG, and the integral parameters $T_{ii}$ and $K_{ii}$ are well defined, with
values of the order of 1 day and 6 $\unit{10^6}{cm^2~sec^{-1}}$ respectively.
In the boundary regions, instead, the along component of the
autocorrelation shows a significant ``tail'' at lags $\tau > $ 4 days,
especially in EAC. This tail can be characterized as a secondary
exponential behavior with slower decay time of $\approx$ 11-12 days. As a
consequence, the integral parameters do not converge for times less than
10 days. Possible physical reasons for this secondary time scale are
discussed, in terms of low frequency fluctuations in the wind regime and
in the Otranto inflow, or in terms of topographic and mean flow
curvatures inducing fluctuations in the particle trajectories. 

The effects of non-stationarity have been partially evaluated
partitioning the data in two extended seasons, corresponding to
winter-spring (January to June) and summer-fall (July-December). The
secondary time scales in the along autocorrelation is found to be
present only during summer-fall, when the mean boundary currents are
more enhanced and more energetic.

On the basis of these statistical analysis, the following indications
for the application of transport models can be given. The statistics of
$\mathbf{u}^\prime$, and therefore its modelling description, are strongly
inhomogeneous in the three regions not only in terms of parameter values
but also in terms of inherent turbulent properties. It is therefore not
surprising that the results of \citet{falco_etal-jpo-2000} show differences
between data and model results, given that the model uses global
parameters and assumes gaussianity over the whole basin. Only region CG can be characterized
by homogeneous and Gaussian turbulence and therefore can be correctly
described using a classical Langevin equation such as the one used by
\citet{falco_etal-jpo-2000}. The boundary regions and especially EAC, on the
other hand, are not correctly described by such a model, because of the
presence of a secondary time scale and of significant deviations from
gaussianity. Similar deviations have been observed in other Lagrangian
data in various ocean regions \citep{bracco_etal-jpo-2000}, even though the
ubiquity of the result is still under debate \citep{zhang_etal-jgr-2001}.
Non-Gaussian, multi scale models are known in the literature
\citep[\eg][]{pasquero_etal-jfm-2001,maurizi_etal-ftc-2002}
and their
application is expected to strongly improve results of transport
modelling in the Adriatic Sea.

\section*{Appendix: Spline method for estimating $\mathbf{U}$}

The spline method used to estimate $\mathbf{U}$
\citep{bauer_etal-jgr-1998,bauer_etal-jgr-2002}
is based on the application of
a bicubic spline interpolation \citep{inoue-geo-1986} with optimized parameters to
guarantee minimum energy of the fluctuation $\mathbf{u}^\prime$ at low frequencies.
This is done
by minimizing a simple metrics which depends on the integration of the
autocovariance $R(\tau)$ for $\tau>T$.
In other words, the autocovariance tail is
required to be ``as flat as possible'' under some additional smoothing
requirements. 
This method, previously applied by Falco et al.
(2000) to the 1994-1996 data set, has been applied
1990-1999 data set.

The spline results depend on four parameters (Inoue, 1986): the values of the
knot spacing, which determines the number of finite elements, and three
weights associated respectively with the uncertainties in the data, in the first
derivatives (tension) and in the second derivatives (roughness). The tension
can be fixed a-priori in order to avoid anomalous behavior
at the boundaries (Inoue, 1986). The other three parameters have been varied
in a wide range of values (knot spacing between $1^{\degree}$ and
$0.1^{\degree}$, data uncertainty between 50 and
$\unit{120}{cm^2sec^{-2}}$ and
roughness between 0.001 and 10000). It is found that an optimal estimate of
$\mathbf{U}$ is uniquely defined over the whole parameter space except for
the smallest knot spacing, corresponding to $0.1^{\degree}$. In this case,
no optimal solution is found, in the sense that the metric does not asymptote
and the $\mathbf{U}$ field becomes increasingly more noisy as the roughness
increases.
This indicates that, as it can be intuitively understood, 
there is a minimum resolution related to the
number of data available. 

\section*{Acknowledgements}

The authors are indebted to Fulvio Giungato for providing helpful
results of a preliminary data analysis performed as a part of his Degree
Thesys at University of Urbino, Italy.
The authors greatly appreciate the support of the SINAPSI Project
(A.Maurizi, A. Griffa, F. Tampieri) and of the Office of Naval Research,
grant (N00014-97-1-0620 to A. Griffa and grants N0001402WR20067 and
N0001402WR20277 to P.-M. Poulain).

\bibliographystyle{personal}
\bibliography{abbr,boe}

\begin{thebibliography}{36}
\expandafter\ifx\csname natexlab\endcsname\relax\def\natexlab#1{#1}\fi

\bibitem[Artegiani et~al., 1997]{artegiani_etal-jpo-1997}
Artegiani, A., D.~Bregant, E.~Paschini, N.~Pinardi, F.~Raicich, and A.~Russo,
  1997: The {Adriatic Sea} general circulation, part {II}: Baroclinic
  circulation structure. \textit{J. Phys. Oceanogr.}, \textbf{27}, 1515--1532.

\bibitem[Batchelor, 1970]{batchelor-1970}
Batchelor, G., 1970: \textit{The theory of homogeneous turbulence}, Cambridge
  University Press.

\bibitem[Bauer et~al., 2002]{bauer_etal-jgr-2002}
Bauer, S., M.~Swenson, and A.~Griffa, 2002: Eddy-mean flow decomposition and
  eddy-diffusivity estimates in the tropical pacific ocean. part 2: Results.
  \textit{J. Geophys. Res.}, \textbf{107}, 3154--3171.

\bibitem[Bauer et~al., 1998]{bauer_etal-jgr-1998}
Bauer, S., M.~Swenson, A.~Griffa, A.~Mariano, and K.~Owens, 1998: Eddy-mean
  flow decomposition and eddy-diffusivity estimates in the tropical pacific
  ocean. part 1: Methodology. \textit{J. Geophys. Res.}, \textbf{103},
  30,855--30,871.

\bibitem[Berloff and McWilliams, 2002]{berloff_etal-jpo-2002}
Berloff, P. and McWilliams, 2002: Material transport in oceanic gyres. part ii:
  Hierarchy of stochastic models. \textit{J. Phys. Oceanogr.}, \textbf{32},
  797--830.

\bibitem[Bracco et~al., 2000]{bracco_etal-jpo-2000}
Bracco, A., J.~LaCasce, and A.~Provenzale, 2000: Velocity pdfs for oceanic
  floats. \textit{J. Phys. Oceanogr.}, \textbf{30}, 461--474.

\bibitem[Castellari et~al., 2001]{castellari_etal-jms-2001}
Castellari, S., A.~Griffa, T.~Ozgokmen, and P.-M. Poulain, 2001: Prediction of
  particle trajectories in the {Adriatic Sea} using lagrangian data
  assimilation. \textit{J. Mar. Sys.}, \textbf{29}, 33--50.

\bibitem[Cerovecki et~al., 1991]{cerovecki_etal-geo-1991}
Cerovecki, I., Z.~Pasaric, M.~Kuzmic, J.~Brana, and M.~Orlic, 1991: Ten-day
  variability of the summer circulation in the {North Adriatic}.
  \textit{Geofizika}, \textbf{8}, 67--81.

\bibitem[Corrsin, 1974]{corrsin-advg-1974}
Corrsin, S., 1974: Limitation of gradient transport models in random walks and
  in turbulence. \textit{Annales Geophysicae}, \textbf{{18A}}, 25--60.

\bibitem[Cushman-Roisin et~al., 2001]{cushman-roisin_etal-2001}
Cushman-Roisin, B., M.~Gacic, P.-M. Poulain, and A.~Artegiani, 2001:
  \textit{Physical oceanography of the {Adriatic Sea}: Past, present and
  future}, Kluwer Academic Publishers, 316 pp.

\bibitem[Davis, 1987]{davis-jmr-1987}
Davis, R., 1987: Modelling eddy transport of passive tracers. \textit{J. Mar.
  Res.}, \textbf{45}, 635--666.

\bibitem[Davis, 1991]{davis-dsr-1991}
Davis, R., 1991: Observing the general circulation with floats.
  \textit{Deep-Sea Research}, \textbf{38}, S531--S571.

\bibitem[Davis, 1994]{davis-1994}
Davis, R., 1994: Lagrangian and eulerian measurements of ocean transport
  processes, \textit{Ocean Processes in climate dynamics: Global and
  Mediterranean examples}, P.~Malanotte-Rizzoli and A.~R. Robinson, eds., Eds.,
  Kluwer Academic Publishers, pp. 29--60.

\bibitem[Durst et~al., 1987]{durst_etal-1987}
Durst, F., J.~Jovanovic, and L.~Kanevce, 1987: Probability density
  distributions in turbulent wall boundary-layer flow, \textit{Turbulent Shear
  Flow 5}, F.~Durst, B.~E. Launder, J.~L. Lumley, F.~W. Schmidt, and J.~H.
  Whitelaw, eds., Springer, pp. 197--220.

\bibitem[Falco et~al., 2000]{falco_etal-jpo-2000}
Falco, P., A.~Griffa, P.-M. Poulain, and A.~Zambianchi, 2000: Transport
  properties in the {Adriatic Sea} as deduced from drifter data. \textit{J.
  Phys. Oceanogr.}, \textbf{30}, 2055--2071.

\bibitem[Griffa, 1996]{griffa-1996}
Griffa, A., 1996: Applications of stochastic particle models to oceanographic
  problems, \textit{Stochastic Modelling in Physical Oceanography}, P.~M.
  R.J.~Adler and B.~Rozovskii, eds., Birkhauser, pp. 114--140.

\bibitem[Inoue, 1986]{inoue-geo-1986}
Inoue, 1986: A least square smooth fitting for irregularly spaced data: Finite
  element approach using the cubic $\beta$-spline. \textit{Geophysics},
  \textbf{51}, 2051--2066.

\bibitem[Lenschow et~al., 1994]{lenschow_etal-jaot-1994}
Lenschow, D.~H., J.~Mann, and L.~Kristensen, 1994: How long is long enough when
  measuring fluxes and other turbulence statistics? \textit{J. Atmos. Ocean.
  Technol.}, \textbf{11}, 661--673.

\bibitem[Maurizi and Lorenzani, 2001]{maurizi_etal-ftc-2002}
Maurizi, A. and S.~Lorenzani, 2001: Lagrangian time scales in inhomogeneous
  non-{Gaussian} turbulence. \textit{Flow, Turbulence and Combustion},
  \textbf{67}, 205--216.

\bibitem[Niiler et~al., 1995]{niiler_etal-dsr-1995}
Niiler, P.~P., A.~S. Sybrandy, K.~Bi, P.-M. Poulain, and D.~S. Bitterman, 1995:
  Measurements of the water following capability of holey- and tristar
  drifters. \textit{Deep-Sea Research}, \textbf{42}, 1951--1964.

\bibitem[Pasquero et~al., 2001]{pasquero_etal-jfm-2001}
Pasquero, C., A.~Provenzale, and A.~Babiano, 2001: Parameterization of
  dispersion in two-dimensional turbulence. \textit{J. Fluid Mech.},
  \textbf{439}, 278--303.

\bibitem[Poulain, 1999]{poulain-jms-1999}
Poulain, P.-M., 1999: Drifter observations of surface circulation in the
  {Adriatic Sea} between december 1994 and march 1996. \textit{J. Mar. Sys.},
  \textbf{20}, 231--253.

\bibitem[Poulain, 2001]{poulain-jms-2001}
Poulain, P.-M., 2001: {Adriatic Sea} surface circulation as derived from
  drifter data between 1990 and 1999. \textit{J. Mar. Sys.}, \textbf{29},
  3--32.

\bibitem[Poulain and Cushman-Roisin, 2001]{poulain_etal-c3-2001}
Poulain, P.-M. and B.~Cushman-Roisin, 2001: {Chap. 3: Circulation},
  \textit{Physical oceanography of the {Adriatic Sea}: Past, present and
  future}, B.~Cushman-Roisin, M.~Gacic, P.-M. Poulain, and A.~Artegiani, eds.,
  Kluwer Academic Publishers, pp. 67--109.

\bibitem[Poulain et~al., 2003]{poulain_etal-2003}
Poulain, P.-M., E.~Mauri, C.~Fayos, L.~Ursella, and P.~Zanasca, 2003:
  Mediterranean surface drifter measurements between 1986 and 1999, Tech. Rep.
  CD-ROM, OGS, in preparation.

\bibitem[Poulain and Raicich, 2001]{poulain_etal-c2-2001}
Poulain, P.-M. and F.~Raicich, 2001: {Chap. 2: Forcings}, \textit{Physical
  oceanography of the {Adriatic Sea}: Past, present and future},
  B.~Cushman-Roisin, M.~Gacic, P.-M. Poulain, and A.~Artegiani, eds., Kluwer
  Academic Publishers, pp. 45--65.

\bibitem[Poulain and Zanasca, 1998]{poulain_etal-1998}
Poulain, P.-M. and P.~Zanasca, 1998: Drifter observations in the {Adriatic Sea}
  (1994-1996) - data report, Tech. Rep. SACLANTCEN Memorandum SM-340, SACLANT
  Undersea Research Centre, La Spezia, Italy, 46 pp.

\bibitem[Press et~al., 1992]{num_recipes}
Press, W.~H., S.~A. Teukolsky, W.~T. Vetterling, and B.~P. Flannery, 1992:
  \textit{Numerical Recipes in {FORTRAN}}, 2nd ed., Cambridge University Press.

\bibitem[Priestley, 1981]{priestly-1981}
Priestley, M., 1981: \textit{Spectral Analysis and Time Series}, Academic
  Press, London, 890 pp.

\bibitem[Raicich, 1996]{raicich-jms-1996}
Raicich, F., 1996: On the fresh water balance of the {Adriatic Sea}. \textit{J.
  Mar. Sys.}, \textbf{9}, 305--319.

\bibitem[Risken, 1989]{risken-1989}
Risken, H., 1989: \textit{The Fokker-Planck Equation: Methods of solutions and
  applications}, Springer-Verlag, Berlin, 472 pp.

\bibitem[Sawford, 1999]{sawford-blm-1999}
Sawford, B.~L., 1999: Rotation of trajectories in {Lagrangian} stochastic
  models of turbulent dispersion. \textit{Boundary-Layer Meteorol.},
  \textbf{93}, 411--424.

\bibitem[Swenson and Niiler, 1996]{swenson_etal-jgr-1996}
Swenson, M. and P.~Niiler, 1996: Statistical analysis of the surface
  circulation of the {California} current. \textit{J. Geophys. Res.},
  \textbf{101}, 22,631.

\bibitem[Thomson, 1987]{thomson-jfm-1987}
Thomson, D., 1987: Criteria for the selection of stochastic models of particle
  trajectories in turbulent flows. \textit{J. Fluid Mech.}, \textbf{180},
  529--556.

\bibitem[Veneziani et~al., 2003]{veneziani_etal-jpo-2003}
Veneziani, M., A.~Griffa, A.~Reynolds, and A.~Mariano, 2003: Oceanic turbulence
  and stochastic models from subsurface lagrangian data for the north-west
  atlantic ocean. \textit{J. Phys. Oceanogr.}, submitted.

\bibitem[Zhang et~al., 2001]{zhang_etal-jgr-2001}
Zhang, H., M.~Prater, and T.~Rossby, 2001: Isopycnal lagrangian statistics from
  the {North Atlantic Current} {RAFOS} floats observations. \textit{J. Geophys.
  Res.}, \textbf{106}, 13,187--13,836.

\end{thebibliography}

\newpage
\thispagestyle{empty}

\section*{Table captions}

\begin{description}
\item[Table 1.] Values of Skewness  $\mathbf{Sk}$ and Kurtosis
$\mathbf{Ku}$  in natural coordinates in the three zones.

\item[Table 2.] Values of correlation time $\mathbf{T}$ and diffusion
coefficient $\mathbf{K}$ in the three zones.

\end{description}

\section*{Figure captions}

\begin{description}
\item[Figure 1.]The Adriatic Sea: a) Topography and drifter deployment
locations; b) Mean flow circulation \citep[Adapted
from][]{poulain-jms-2001}.

\item[Figure 2.] Binned Eddy Kinetic Energy (EKE) and Mean Kinetic
Energy (MKE) computed over the whole basin versus bin size $\ur{L}{a}$.
Also indicated are EKE from spline estimates and number of independent
data used in the estimates, as ratio between data belonging to
significant bins, $N_{\ur{L}{a}}$, and total amount of data,
$\ur{N}{tot}$.

\item[Figure 3.] Ratio $\sqrt{\mbox{EKE}/\mbox{MKE}}$ versus
$\sqrt{\mbox{MKE}}$ for significant $0.25^{\degree} \times 0.25^{\degree}$ bins in the
basin.

\item[Figure 4.] Mean flow and homogeneous regions: EAC (green), WAC
(red), CG (blue), NR (black).
 
\item[Figure 5.] Binned a) Skewness and b) Kurtosis in Cartesian
coordinates ($x\equiv$ zonal, $y\equiv$ meridional) computed over the whole basin versus bin size $\ur{L}{a}$.
Also indicated is the number of independent data used in the estimates,
as ratio between data belonging to significant bins, $N_{\ur{L}{a}}$,
and total amount of data, $\ur{N}{tot}$.
 
\item[Figure 6.] Pdfs of turbulent velocity $\mathbf{u}^\prime$ in
natural coordinates for the three regions: a) EAC; b) WAC; c) CG 

\item[Figure 7.] Autocorrelations $\mathbf{\rho}$ of turbulent velocity
(logarithmic scale)
$\mathbf{u}^\prime$ in natural coordinates for the three regions and for
the whole basin: a) along component $\rho_{\parallel}$; b) cross
component $\rho_{\perp}$. Results are presented with symbols and model
fits with solid lines.

\item[Figure 8.] Integral time scales $\mathbf{T}$ of turbulent
velocity $\mathbf{u}^\prime$ in natural coordinates for the three
regions: a) along component $T_{\parallel}$; b) cross component
$T_{\perp}$

\item[Figure 9.] Seasonal mean flow: a) winter-spring season; b)
summer-fall season 
 
\item[Figure 10.] Autocorrelations $\mathbf{\rho}$ of turbulent velocity
(logarithmic value)
$\mathbf{u}^\prime$ in natural coordinates for the 2 extended seasons
computed over the whole basin: a) along component $\rho_{\parallel}$; b)
cross component $\rho_{\perp}$

\end{description}

\pagestyle{empty}
\newpage

\begin{table}
\begin{tabular}{c|cc|cc}
zone&$\Sk_{\parallel}$&$\Sk_{\perp}$&$\Ku_{\parallel}$&$\Ku_{\perp}$\\
\hline
EAC&0.48&-0.14&3.9&4.1\\
CG&0.16&-0.02&4.1&4.2\\
WAC&0.52&0.09&3.8&4.1\\

\end{tabular}
\caption{ }
\label{tab:1}
\end{table}

\begin{table}
\begin{tabular}{c|cc|cc}
zone&$T_{\parallel}$&$T_{\perp}$&$K_{\parallel}$&$K_{\perp}$\\
\hline
EAC&2.7&.78&38$\times 10^6$&3.1$\times 10^6$\\
CG&1.2&.63&6.9$\times 10^6$&2.9$\times 10^6$\\
WAC&2.0&.52&29$\times 10^6$&1.4$\times 10^6$\\

\end{tabular}
\caption{ }
\label{tab:2}
\end{table}

\newpage
\clearpage

\clearpage
\addtocounter{figure}{-1}
\begin{figure}
\includegraphics[width=\textwidth]{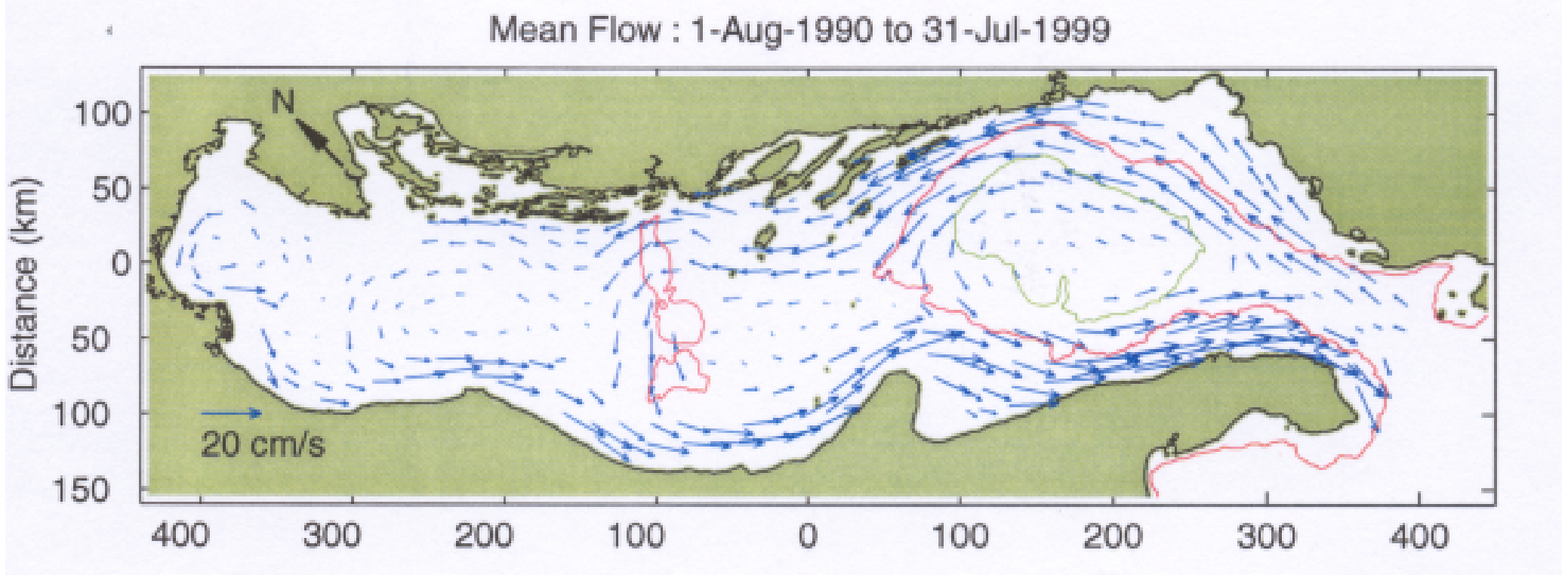}
\caption{}
\label{fig:mean}
\end{figure}
\clearpage

\begin{figure}
\includegraphics[width=\textwidth]{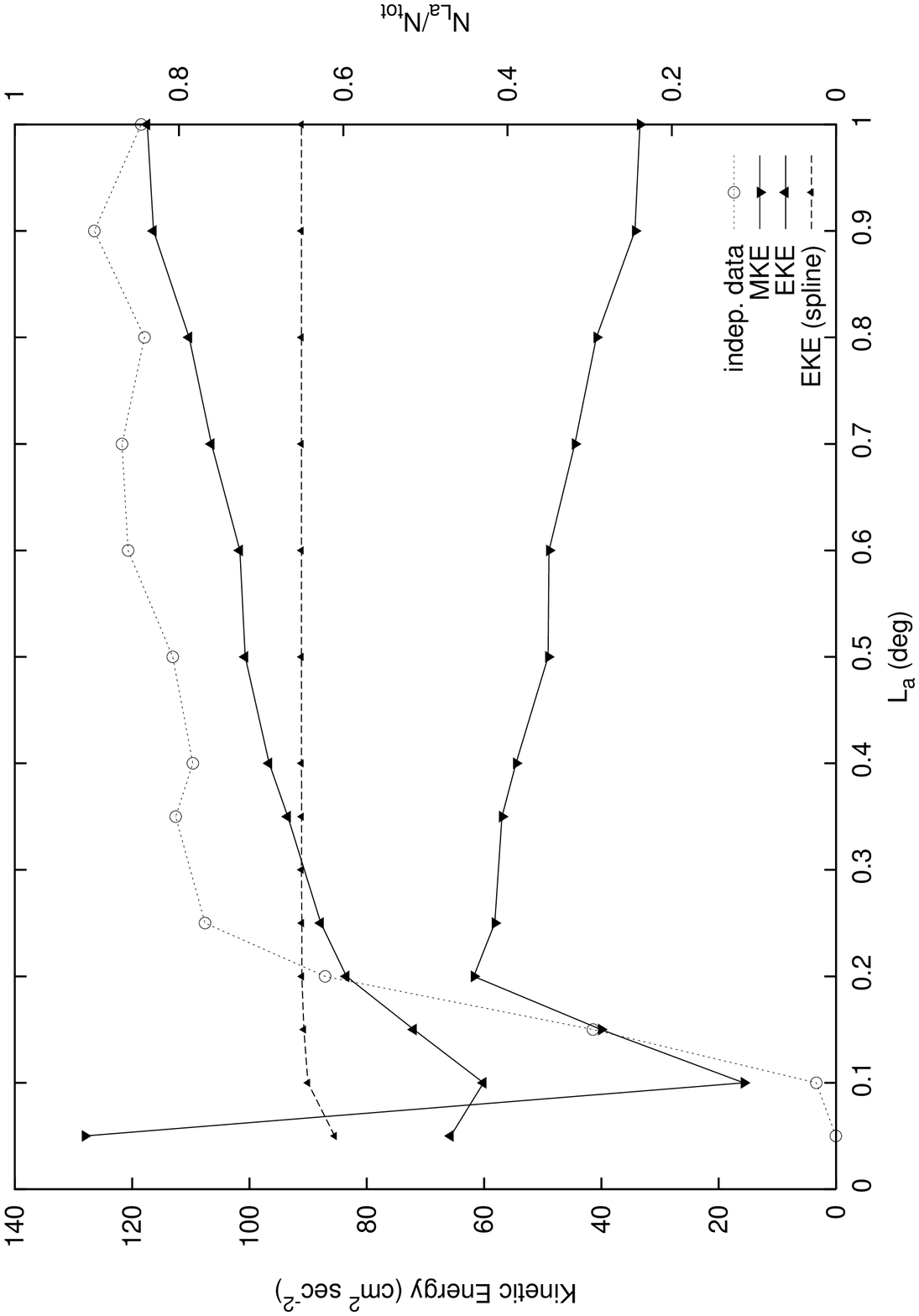}
\caption{}
\label{fig:binning}
\end{figure}
\clearpage

\begin{figure}
\includegraphics[width=\textwidth]{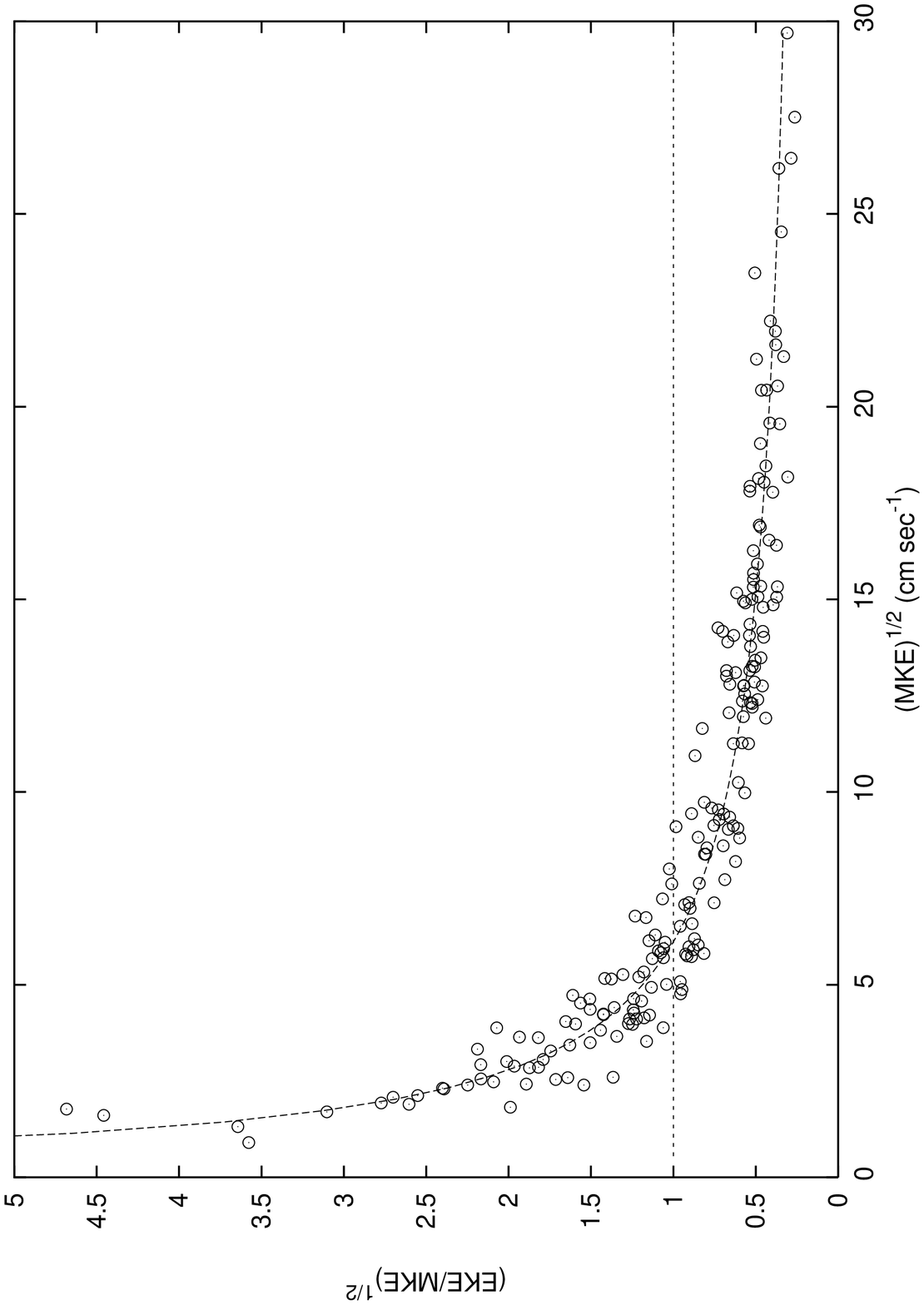}
\caption{}
\label{fig:partition}
\end{figure}
\clearpage

\begin{figure}
\includegraphics[width=\textwidth]{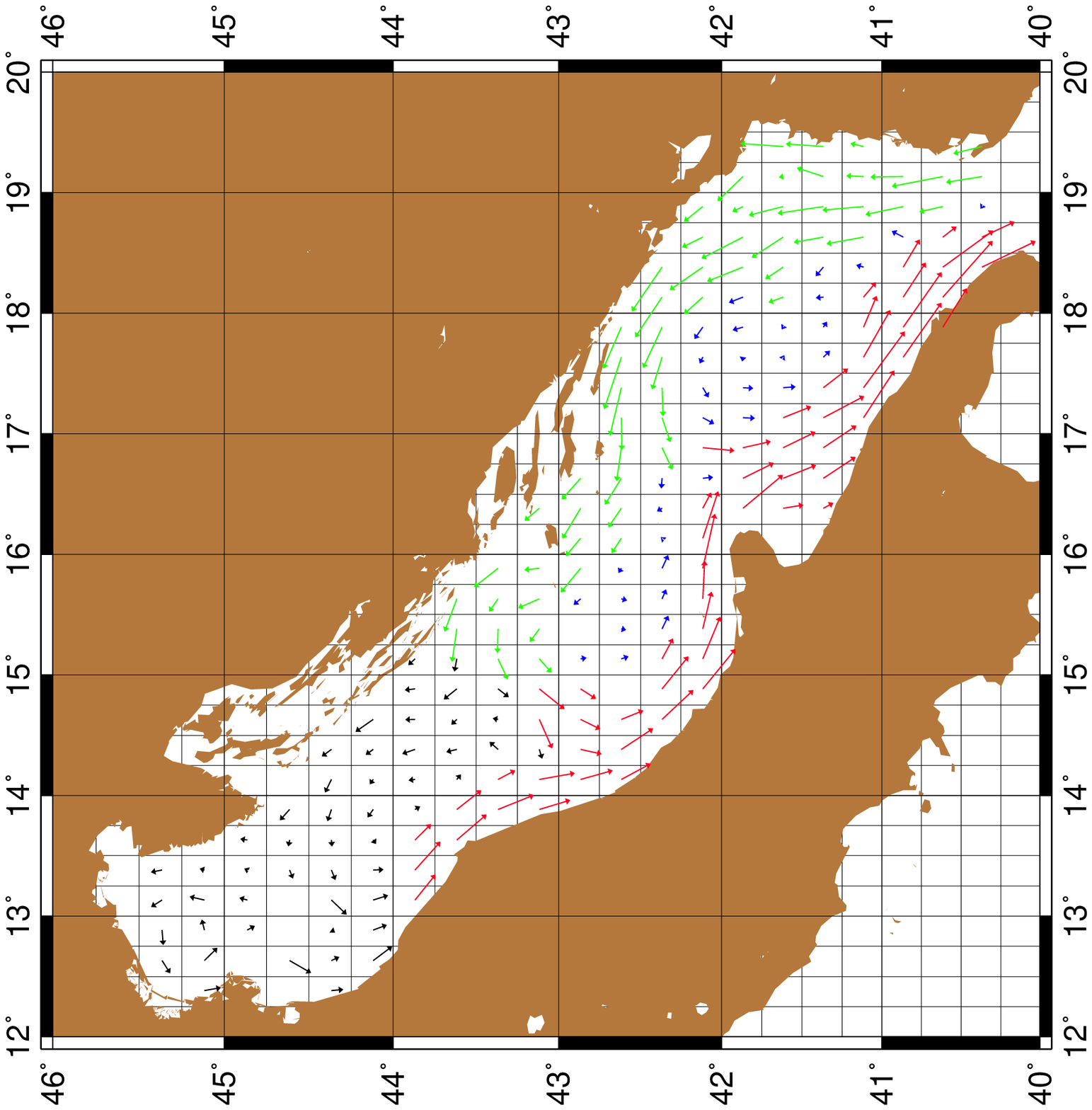}
\caption{}
\label{fig:mean-flow}
\end{figure}
\clearpage

\begin{figure}
\includegraphics[width=\textwidth]{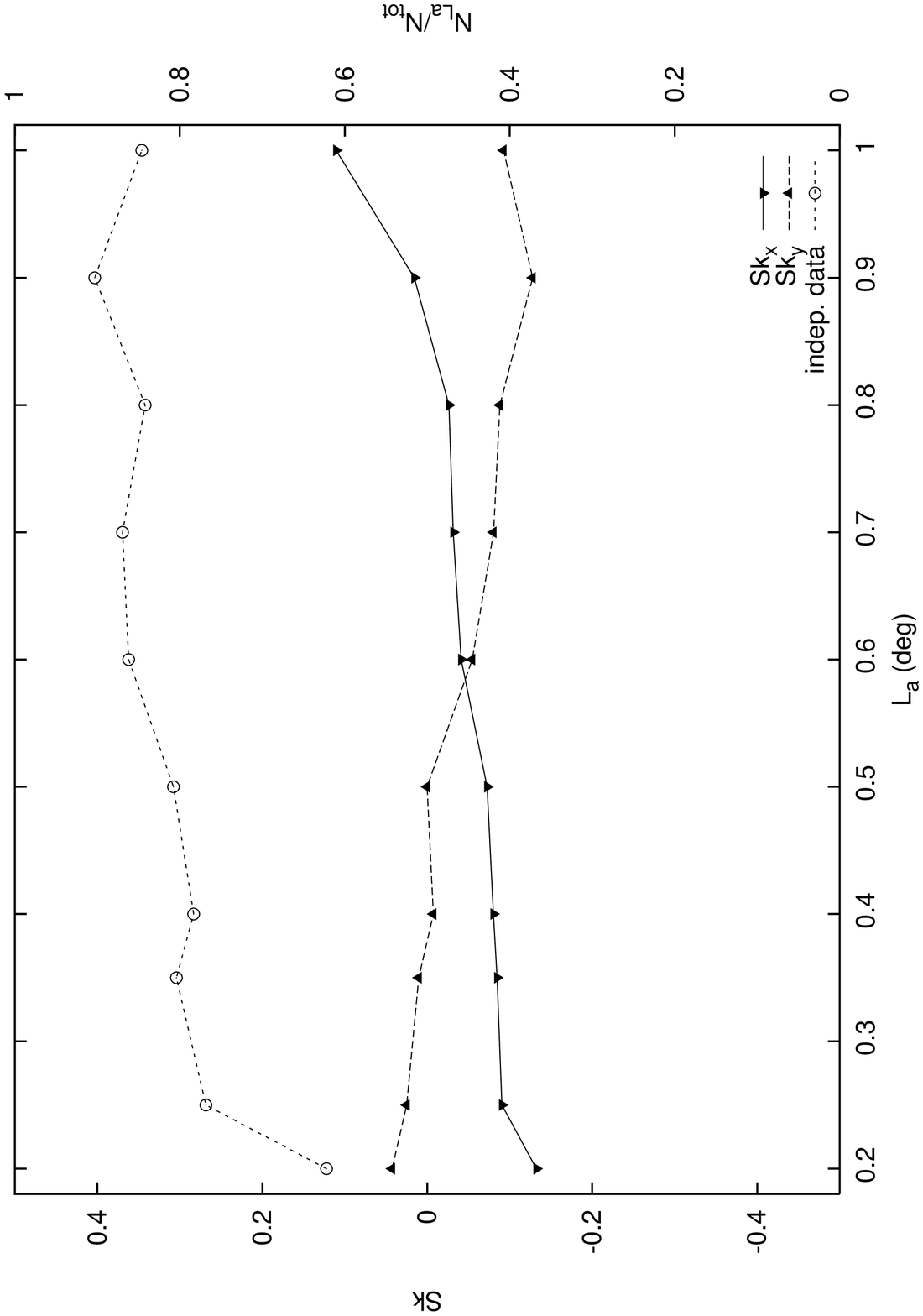}
\caption{}
\label{fig:binning-S}
\end{figure}
\clearpage
\addtocounter{figure}{-1}
\begin{figure}
\includegraphics[width=\textwidth]{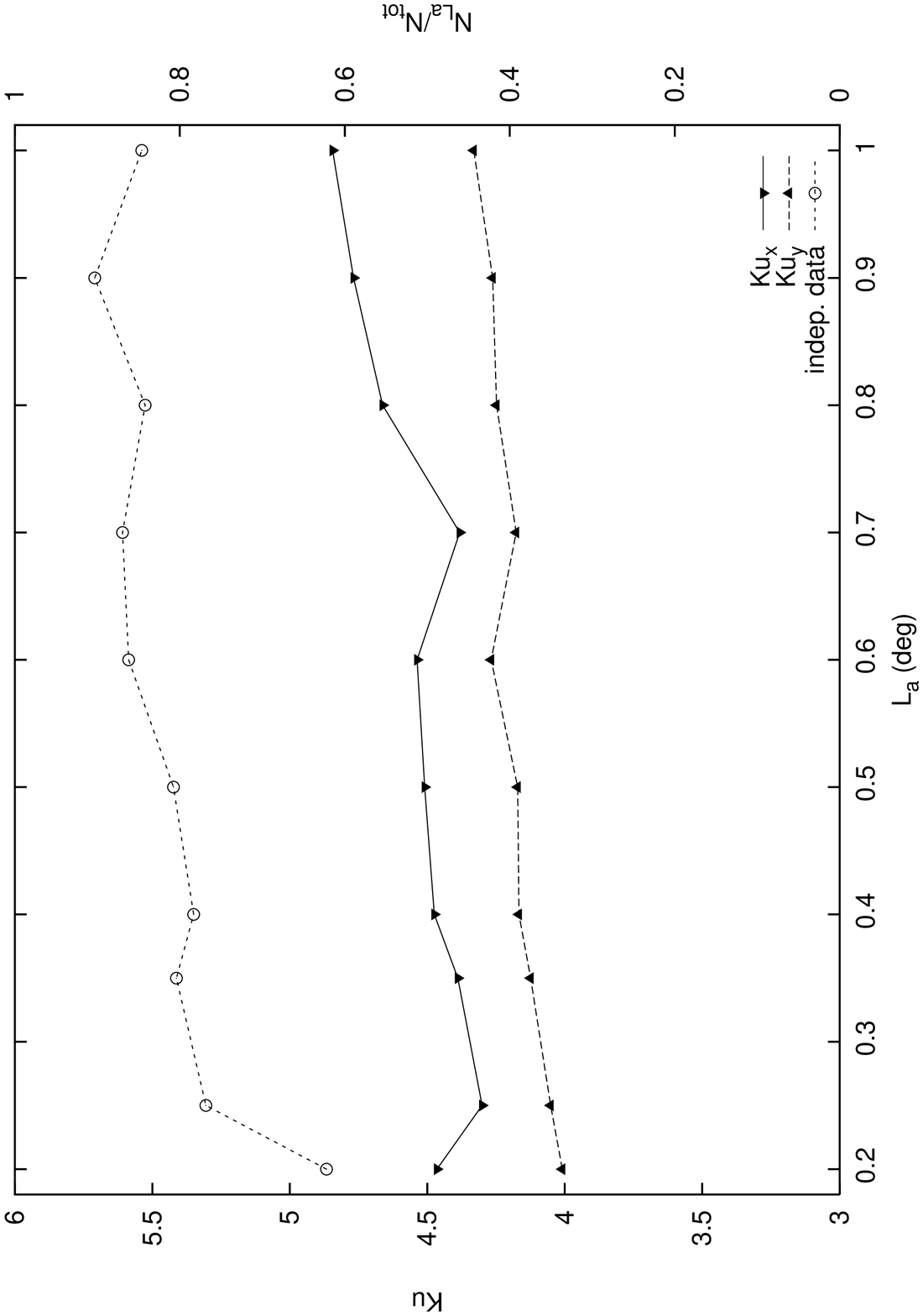}
\caption{}
\label{fig:binning-K}
\end{figure}
\clearpage

\begin{figure}
\includegraphics[width=\textwidth]{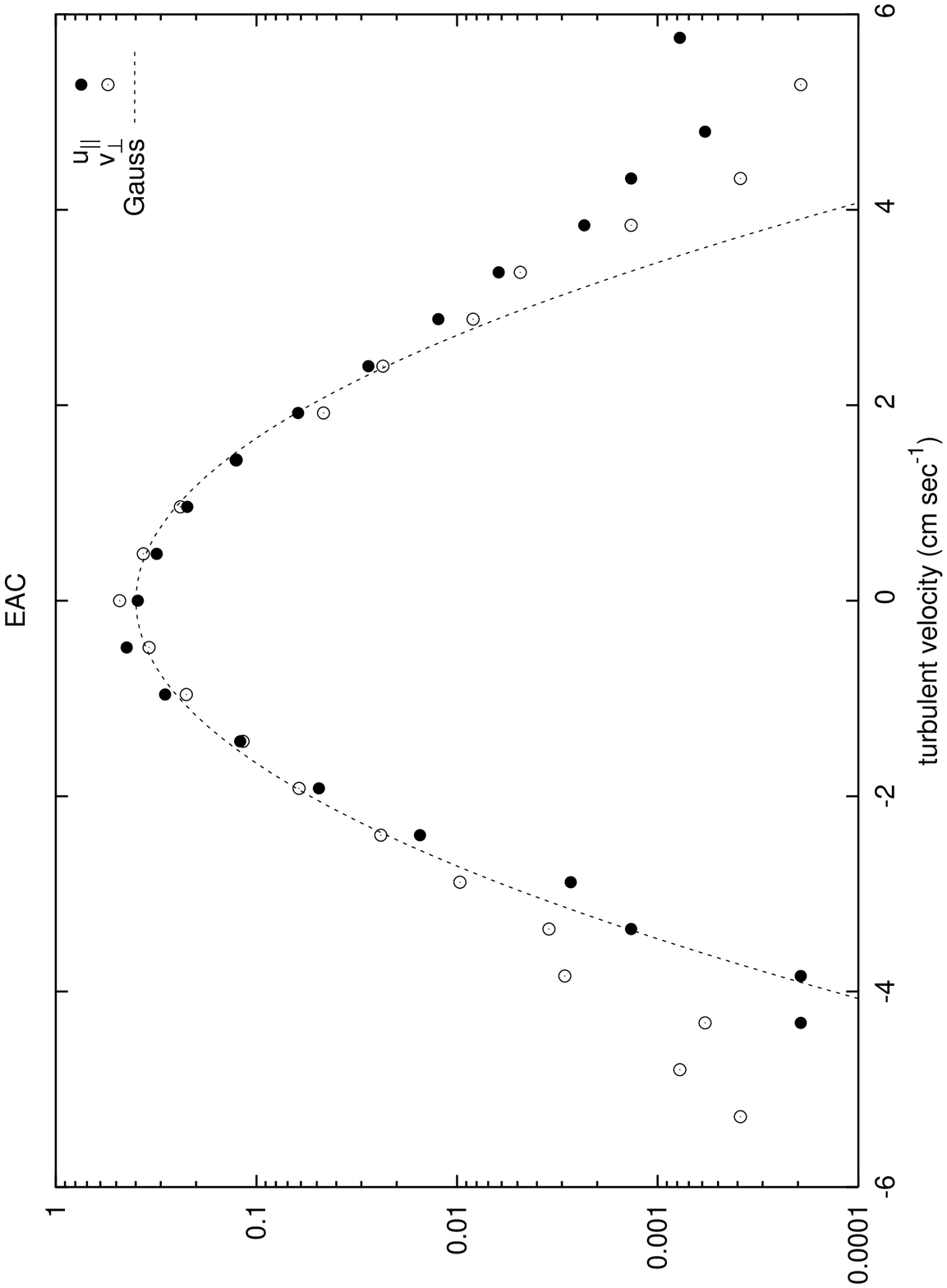}
\caption{}
\label{fig:pdf-EAC}
\end{figure}
\clearpage
\addtocounter{figure}{-1}
\begin{figure}
\includegraphics[width=\textwidth]{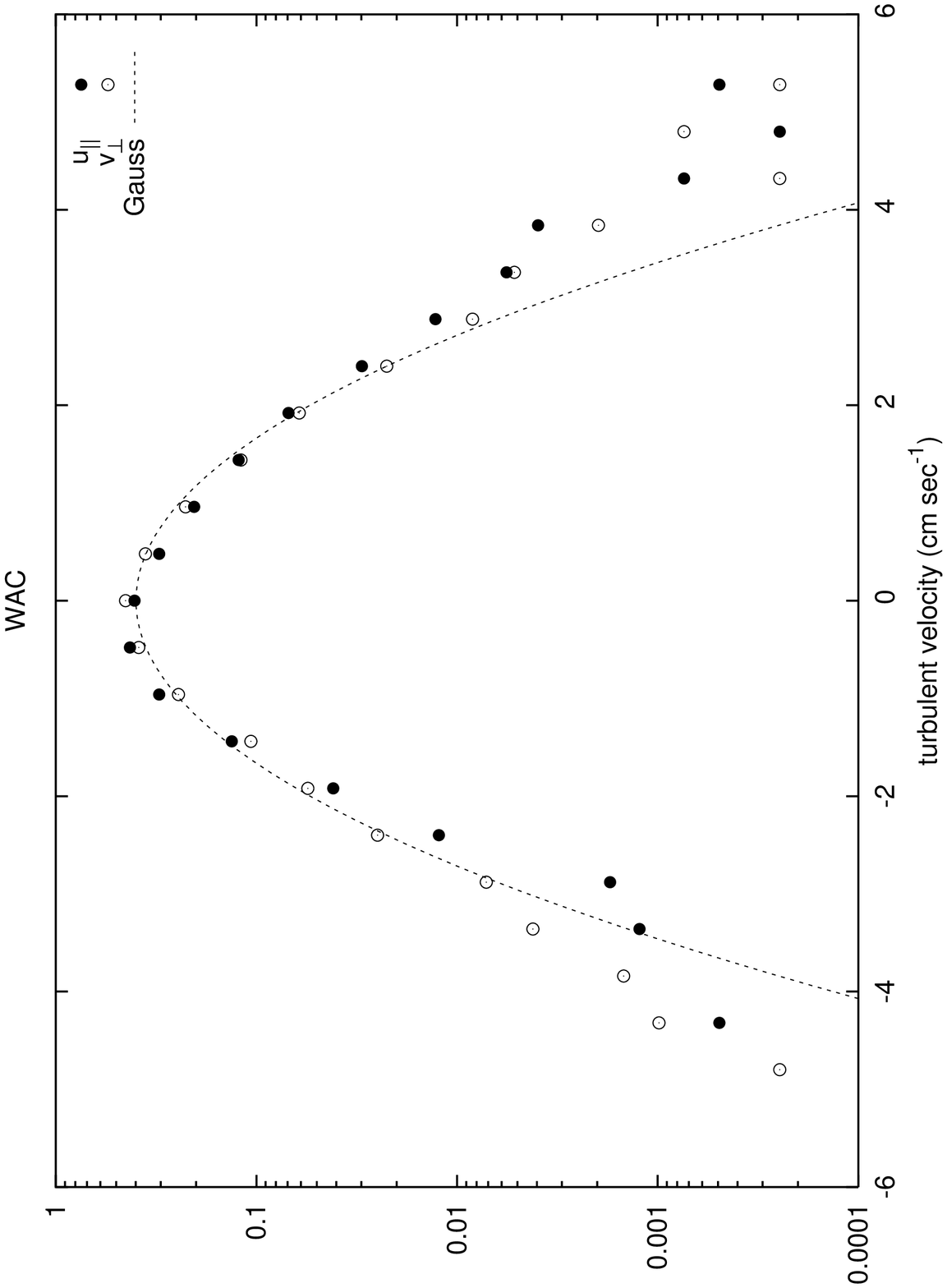}
\caption{}
\label{fig:pdf-WAC}
\end{figure}
\clearpage
\addtocounter{figure}{-1}
\begin{figure}
\includegraphics[width=\textwidth]{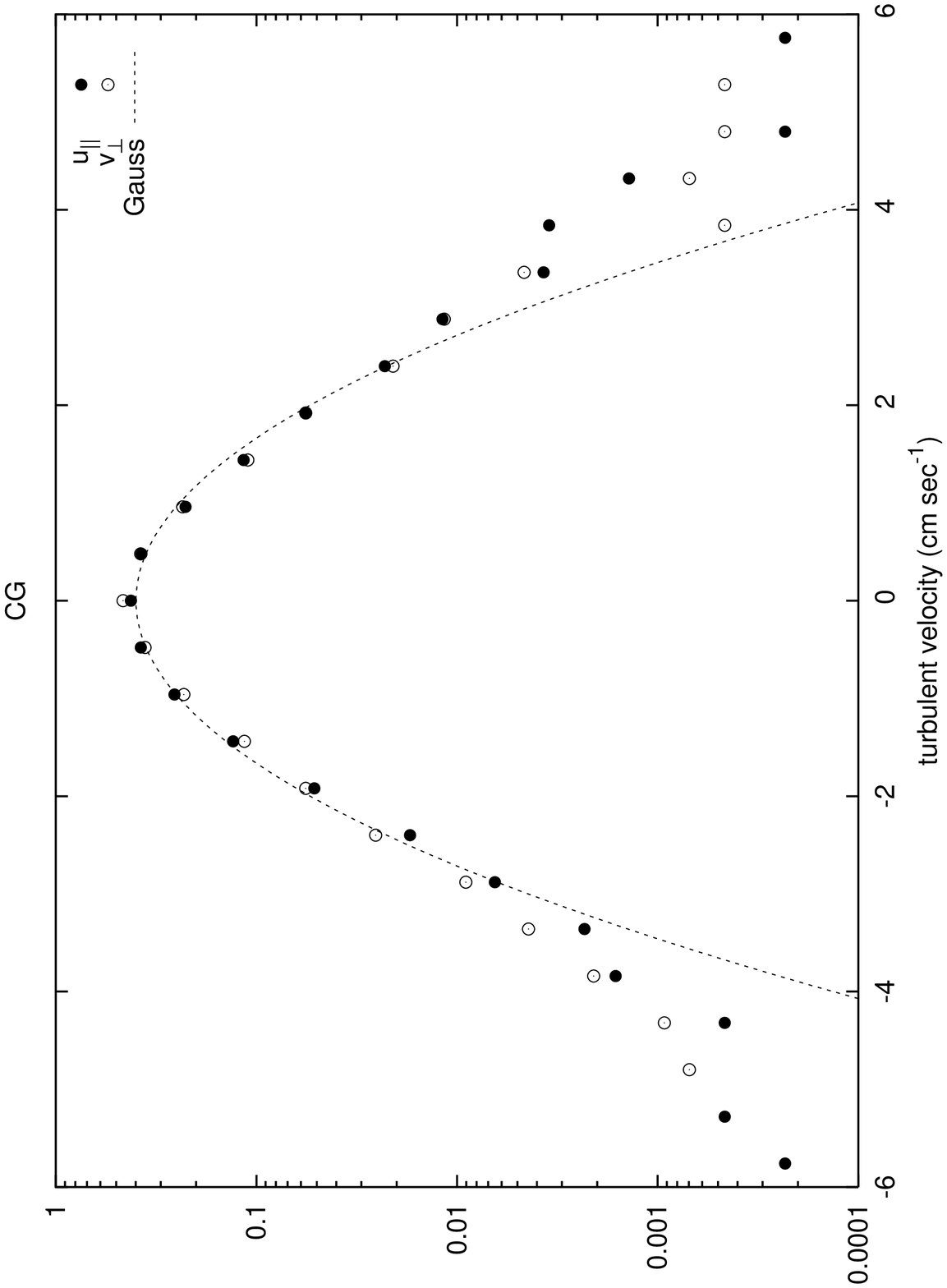}
\caption{}
\label{fig:pdf-CG}
\end{figure}
\clearpage

\begin{figure}
\includegraphics[width=\textwidth]{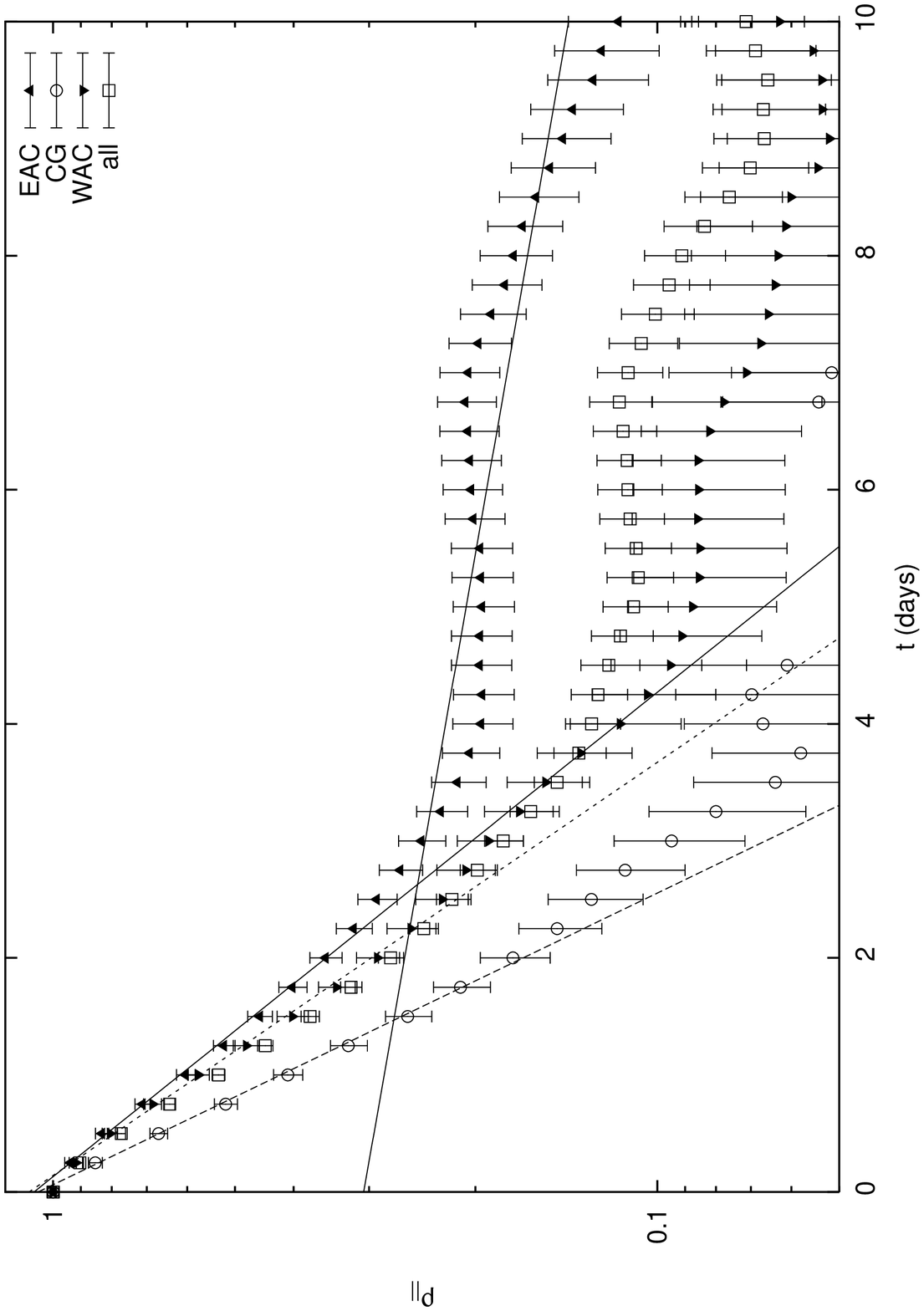}
\caption{}
\label{fig:corr-u}
\end{figure}
\clearpage
\addtocounter{figure}{-1}
\begin{figure}
\includegraphics[width=\textwidth]{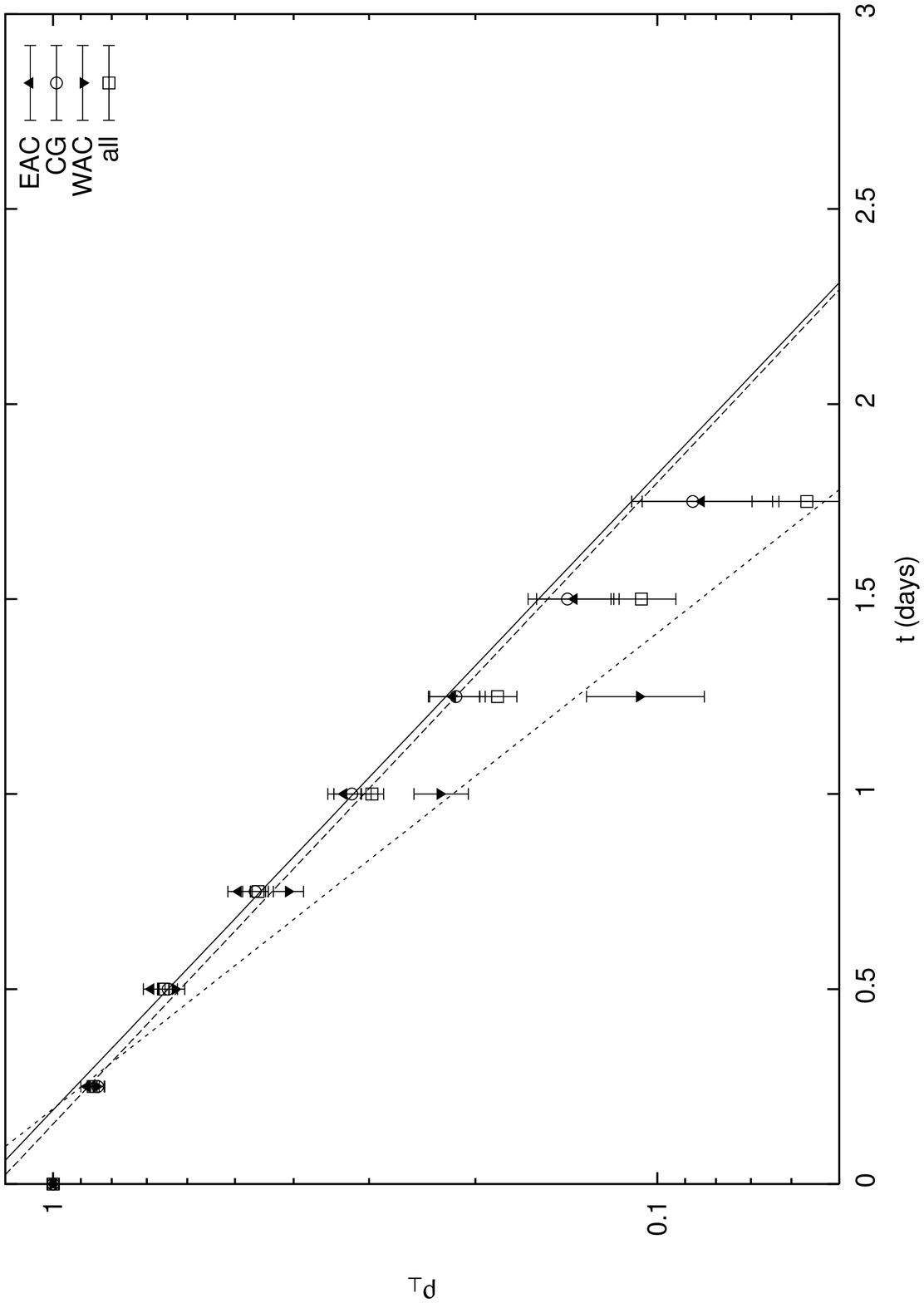}
\caption{}
\label{fig:corr-v}
\end{figure}
\clearpage

\begin{figure}
\includegraphics[width=\textwidth]{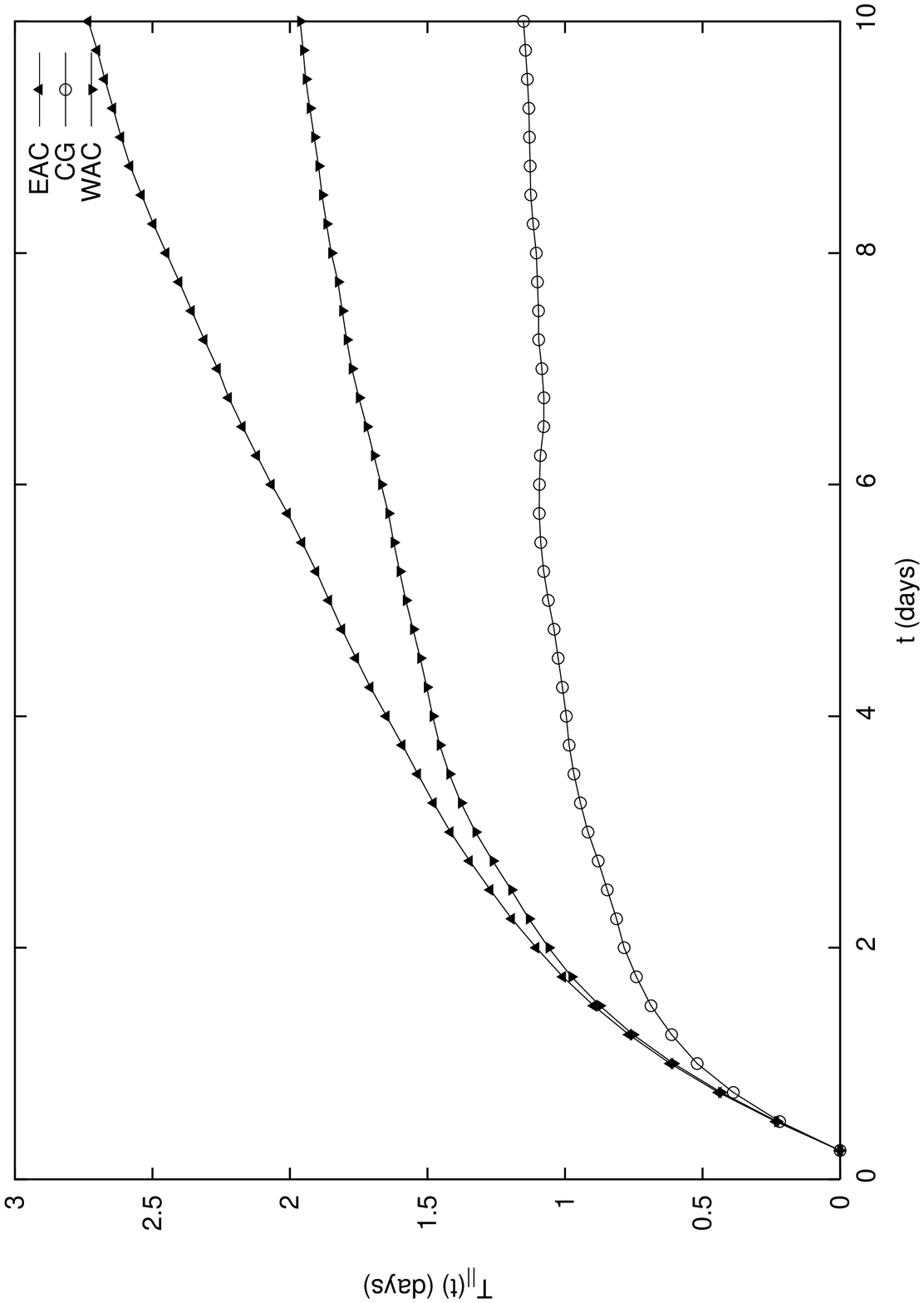}
\caption{}
\label{fig:T-u}
\end{figure}
\clearpage
\addtocounter{figure}{-1}
\begin{figure}
\includegraphics[width=\textwidth]{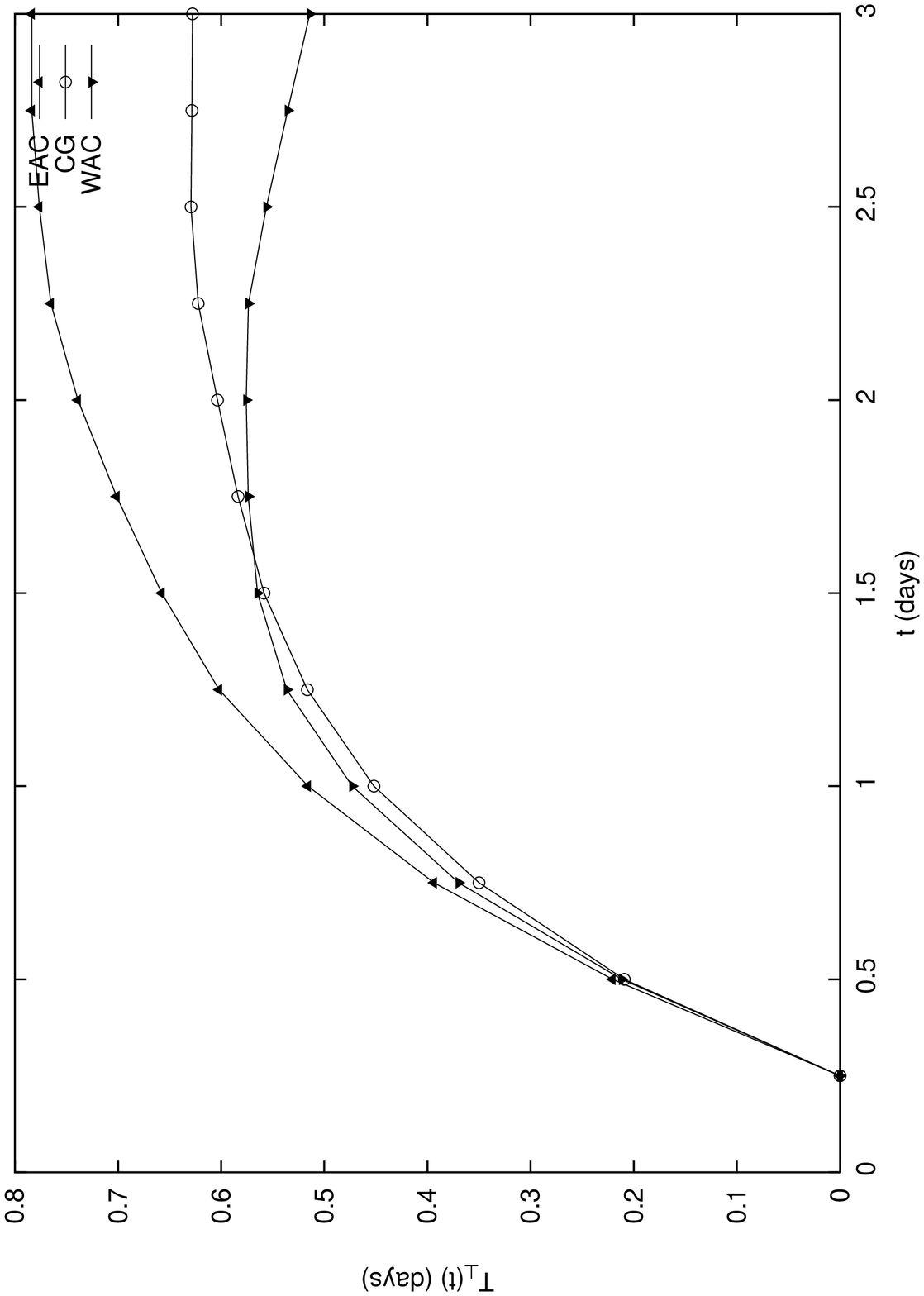}
\caption{}
\label{fig:T-v}
\end{figure}
\clearpage

\begin{figure}
\includegraphics[width=\textwidth]{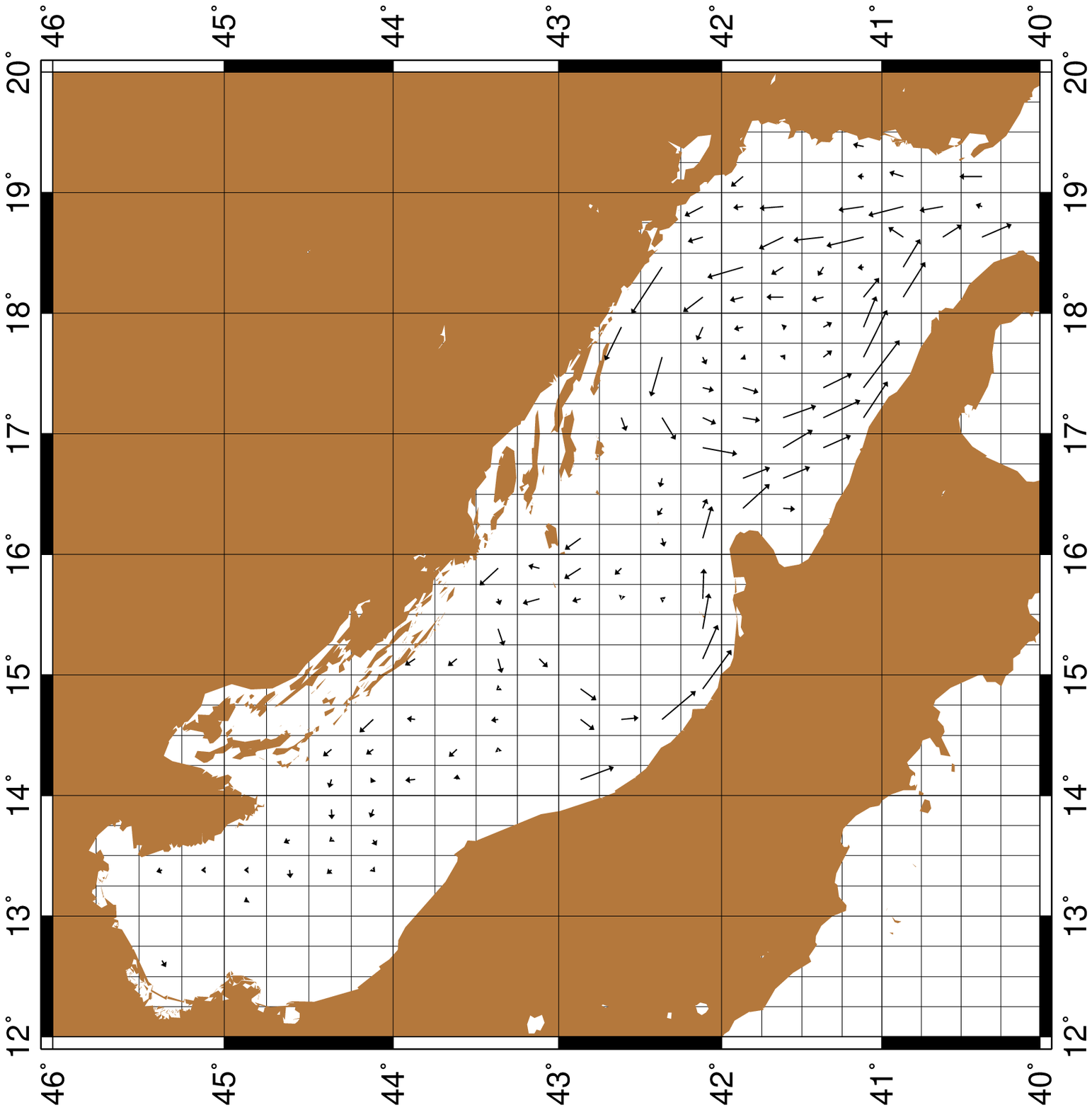}
\caption{}
\label{fig:mean-season-u}
\end{figure}
\clearpage
\addtocounter{figure}{-1}
\begin{figure}
\includegraphics[width=\textwidth]{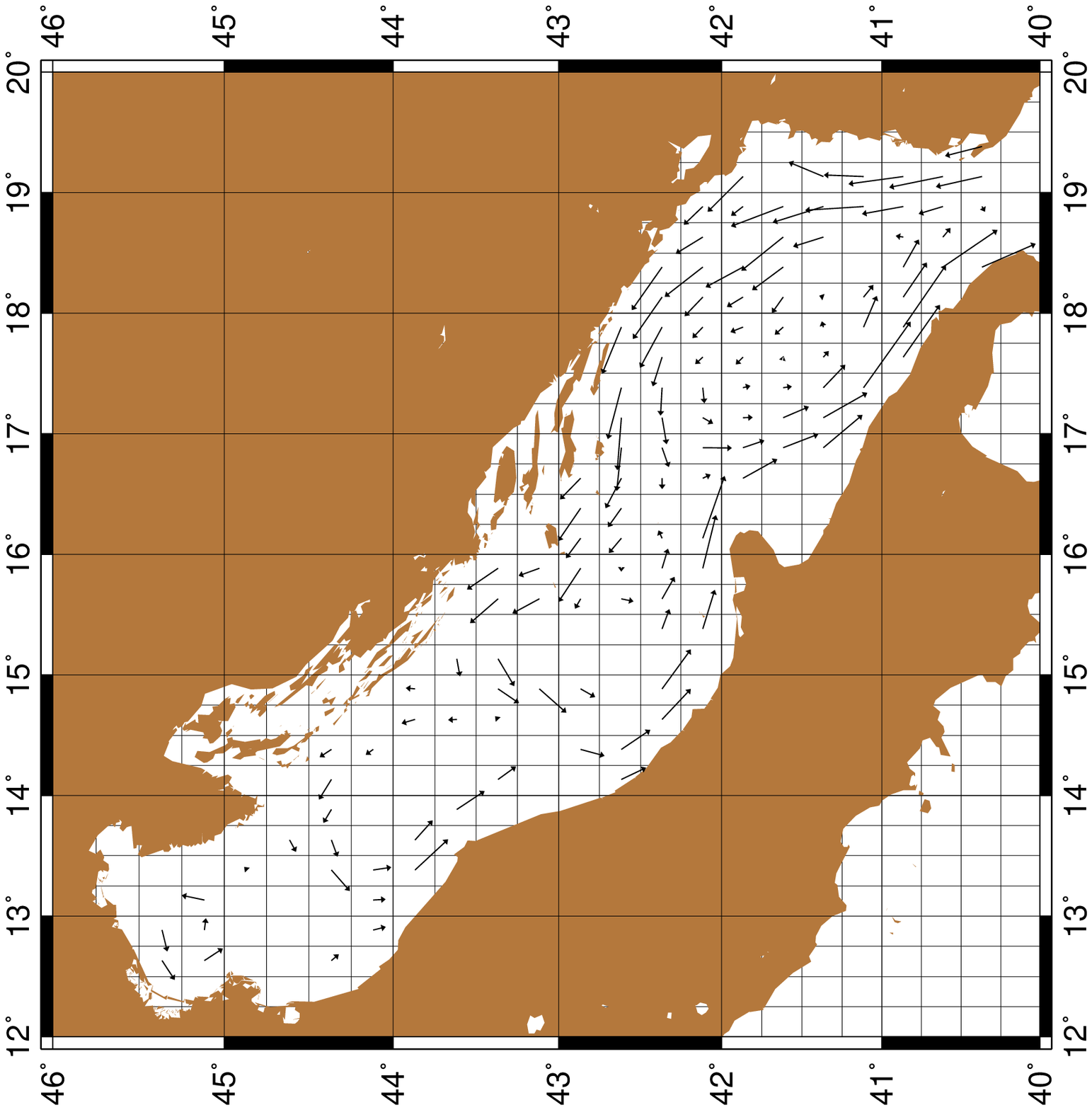}
\caption{}
\label{fig:mean-season-v}
\end{figure}
\clearpage

\begin{figure}
\includegraphics[width=\textwidth]{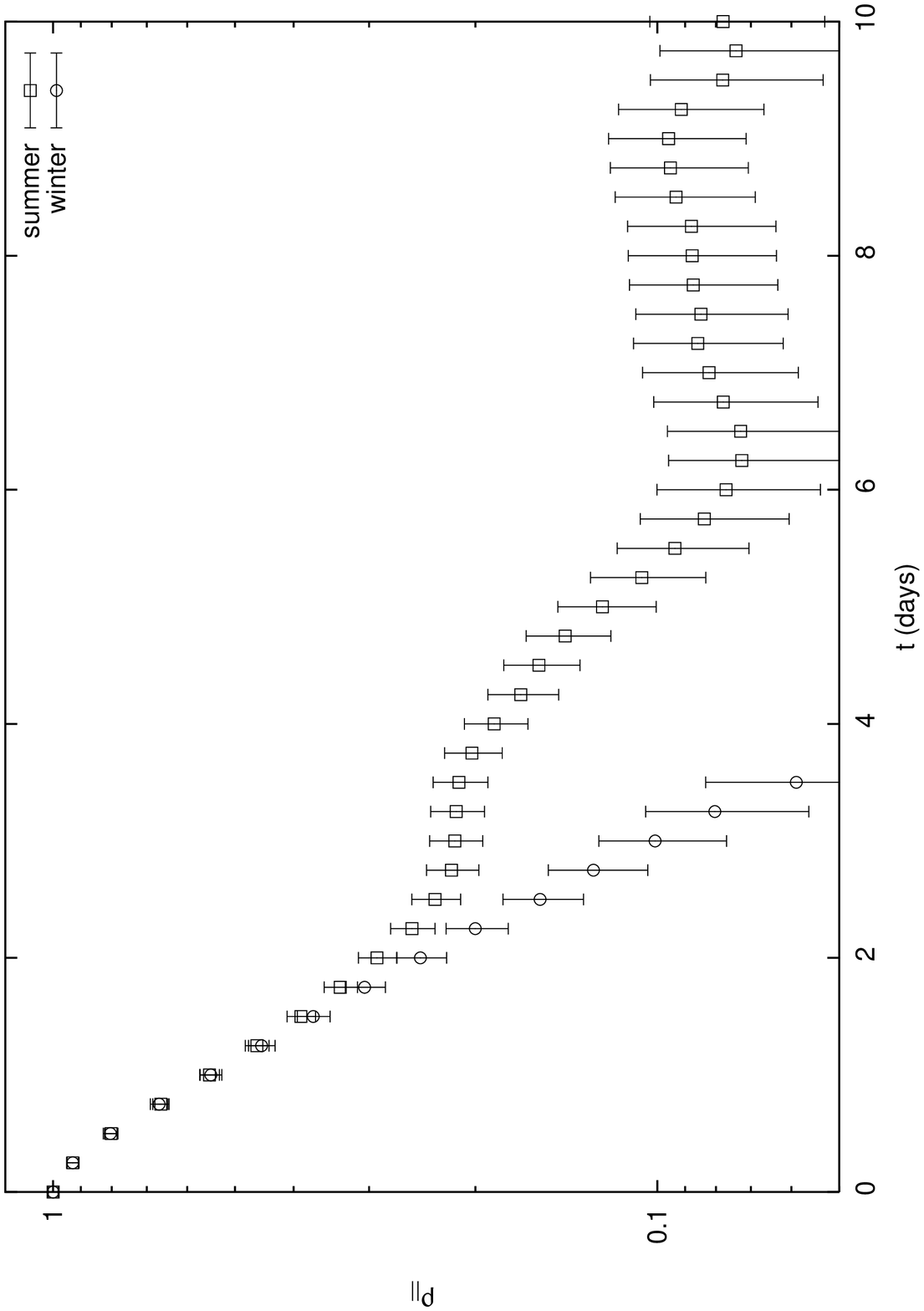}
\caption{}
\label{fig:corr-season-u}
\end{figure}
\clearpage
\addtocounter{figure}{-1}
\begin{figure}
\includegraphics[width=\textwidth]{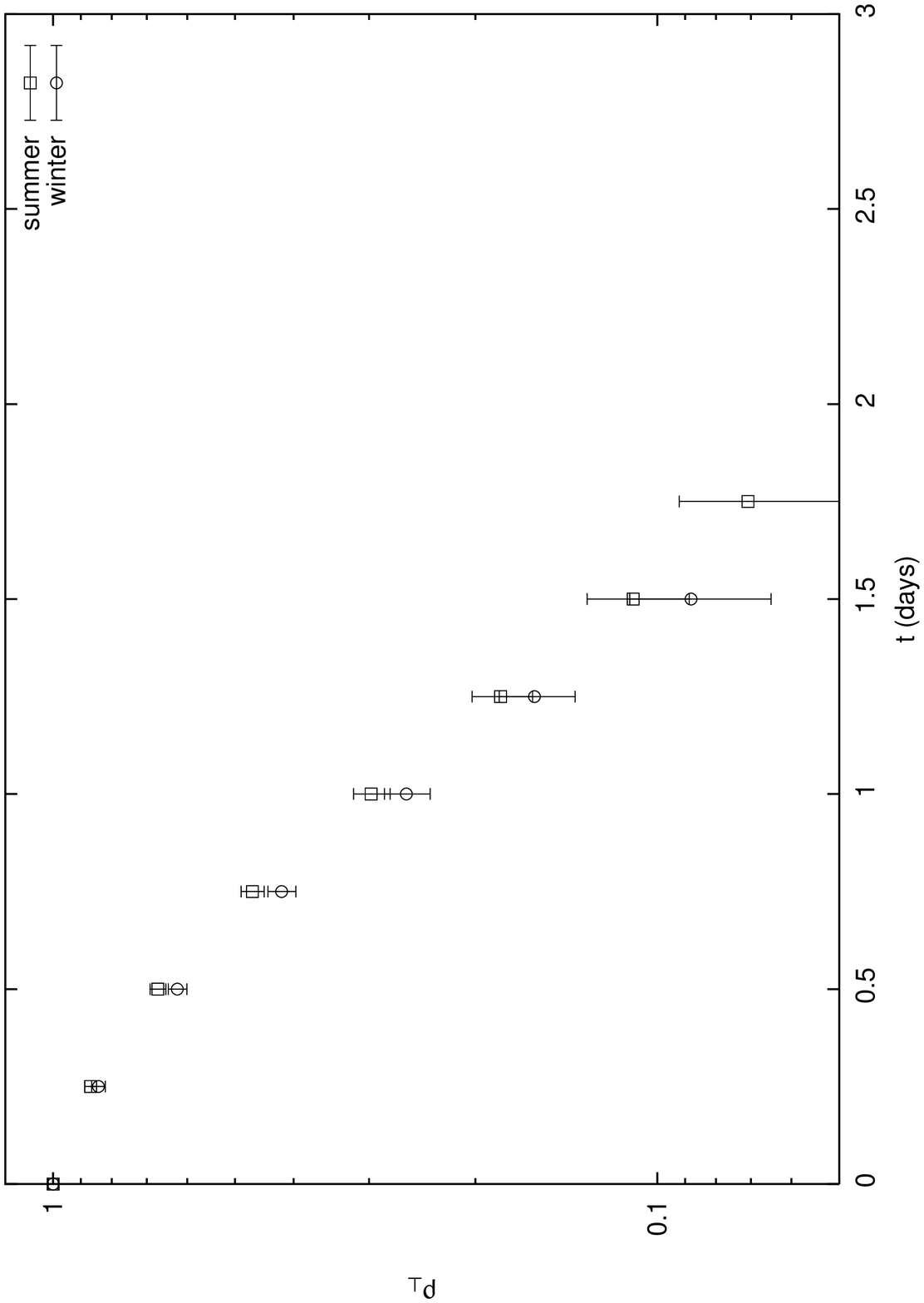}
\caption{}
\label{fig:corr-season-v}
\end{figure}
\clearpage

\end{document}